\documentclass[%
 reprint,
 amsmath,
 amssymb,
 aps,
 prb,
longbibliography,
]{revtex4-1}

\usepackage{graphicx}
\usepackage{dcolumn}
\usepackage{bm}

\usepackage{physics}
\usepackage{color}

\newcommand{\fr}{\mathrm{(FR)}}
\newcommand{\sr}{\mathrm{(SR)}}
\newcommand{\so}{\mathrm{(SO)}}
\newcommand{\p}{\mathrm{(\mathbf{p})}}
\newcommand{\vecop}[1]{\hat{\mathbf{#1}}}

\begin{document}

\title{Effects of spin-orbit coupling on the optical response of a material}

\author{Tae Yun Kim$^1$}
\author{Andrea Ferretti$^2$}
\author{Cheol-Hwan Park$^1$}
\email{cheolhwan@snu.ac.kr}
\affiliation{
$^1$Department of Physics, Seoul National University, Seoul 08826, Korea\\
$^2$Centro S3, CNR-Istituto Nanoscienze, 41125 Modena, Italy
}

\date{\today}

\begin{abstract}
We investigate the effects of 
spin-orbit coupling on the optical response of materials.
In particular,
we study 
the effects of the commutator
between the spin-orbit coupling part of the potential
and the position operator
on the optical matrix elements.
Using a formalism that separates 
a fully-relativistic Kleinman-Bylander pseudopotential 
into the scalar-relativistic and spin-orbit-coupling parts, 
we calculate 
the contribution
of the commutator 
arising from spin-orbit coupling
to the squared optical matrix elements of 
isolated atoms, 
monolayer transition metal dichalcogenides,
and topological insulators.
In the case of isolated atoms
from H ($Z=1$) to Bi ($Z=83$),
the contribution of spin-orbit coupling
to the squared matrix elements
can be as large as 14~\%.
On the other hand,
in the cases of
monolayer transition metal dichalcogenides
and topological insulators,
we find that
this contribution is less than 1~\% and
that 
it is sufficient to calculate 
the optical matrix elements
and subsequent physical quantities 
without considering 
the commutator arising from spin-orbit coupling.
\end{abstract}

\maketitle


\section{Introduction}
The electronic structure of materials 
containing heavy elements can be 
significantly affected by spin-orbit coupling (SOC).
Due to the recent advances 
in the investigation of materials 
having strong SOC effects 
such as transition metal dichalcogenides (TMDCs), 
topological insulators, or
Weyl semimetals to name a few,
it becomes important to accurately simulate 
the effects of SOC using 
first-principles 
density functional theory (DFT) calculations.
Because SOC allows
the manipulation of 
the spin degrees of freedom in materials 
by using light,
\cite{jozwiak2013photoelectron, park2012spin, ryoo2016spin, ryoo2018spin, PhysRevLett.110.216401, PhysRevLett.112.076802} 
understanding the effects of SOC on the optical response of materials is a matter of importance.

Consider a system described 
by an effective Hamiltonian
\begin{equation}
\label{eqn:H}
\hat{H} =
\frac{\vecop{p}^2}{2m} + 
V_{\mathrm{loc}}(\vecop{r}) + 
\hat{V}_{\mathrm{NL}},
\end{equation}
where $m$ is the mass of an electron, 
$\vecop{p}$ is the momentum operator, 
$\vecop{r}$ is the position operator, 
and $V_{\mathrm{loc}}(\vecop{r})$ and 
$\hat{V}_{\mathrm{NL}}$ are 
the local and non-local parts of the potential, 
respectively.
The optical matrix elements of the system 
are given by the matrix elements of 
the velocity operator~\cite{PhysRevB.53.9797}
\begin{equation}
\label{eq_v}
\hat{\mathbf v}
= \frac{\hat{\mathbf{p}}}{m} + \frac{i}{\hbar} \, \left[\hat{V}_{\mathrm{NL}} , \hat{\mathbf{r}} \right].
\end{equation}

In many DFT calculations, 
the pseudopotential method is used 
because of its computational efficiency.
Within the non-relativistic and 
scalar-relativistic pseudopotential methods,
the effects of the commutator in Eq.~(\ref{eq_v})
on the optical matrix elements and absorption spectra
have been investigated 
for various types of systems such as 
isolated atoms,~\cite{PhysRevB.44.13071} 
semiconductors,~\cite{PhysRevB.33.7017} 
and metals.~\cite{PhysRevB.64.195125}
It was reported that 
the contribution of the commutator 
can be large, e.g. as in the cases of 
carbon atom~\cite{PhysRevB.44.13071} and 
bulk copper.~\cite{PhysRevB.64.195125}

SOC is proportional to 
$\vecop{L} \cdot \vecop{S}$, 
where $\vecop{L}$ and $\vecop{S}$ 
are the orbital and spin angular momentum operators, respectively.
Because the orbital angular momentum operator 
does not commute with the position operator, 
SOC results in an additional contribution to 
the velocity operator 
via the commutator
and to the optical matrix elements.

It has not been well established 
whether the effects of the commutator
arising from the SOC part of the potential
are important or not
in a system 
where the influence of SOC 
on the electronic structure
is known to be strong.
For example, 
in some previous studies 
on the optical response of Bi$_2$Se$_3$,
a topological insulator 
that has been extensively investigated,
the contribution of the commutator arising from SOC
was neglected and 
$\vecop{p}/m$ 
as an approximation to $\vecop{v}$ [Eq.~(\ref{eq_v})]
was used to calculate 
the optical matrix elements.~\cite{PhysRevLett.110.216401, PhysRevLett.112.076802, Cao2013}
On the other hand, 
the authors of a recent study~\cite{PhysRevLett.115.016801} 
on the circular dichroism of Bi$_2$Te$_3$ claimed 
that  
the SOC contribution to the velocity operator
plays a crucial role in explaining 
the results of their photoemission experiments.

In this study, 
we investigate the effects of SOC
on the optical matrix elements and absorption spectra
in various types of systems: 
isolated atoms, monolayer TMDCs, and topological insulators.
The method used in this study allows for
the calculation of the optical matrix elements 
with and without 
inclusion of the commutator 
arising from the intrinsic non-locality of SOC 
while using the same (fully-relativistic) pseudopotential,
from which we can directly assess 
the importance of 
the effects of SOC 
in evaluating
the optical matrix elements and 
optical properties of materials.

\section{Methods}

The non-local part of 
a fully-relativistic pseudopotential
in semi-local form
can be written as~\cite{PhysRevB.21.2630}
\begin{equation}
\label{eqn:Vsl}
\begin{aligned}
\hat{V}_{\mathrm{SL}} =
\sum_{l = 0}^{l_{\mathrm{max}}} 
\sum_{j = \left|l - \frac{1}{2} \right|}^{l + \frac{1}{2}} 
\sum_{m_j = -j}^{j}
\ket{l, j, m_j}  V_{l, j}(r)  \bra{l, j, m_j},
\end{aligned}
\end{equation}
where $V_{l, j}(r)$ is 
the radial potential of $\hat{V}_{\mathrm{SL}}$ 
for a given pair of 
the orbital angular momentum quantum number $l$ 
and the total angular momentum $j$
and $\ket{l, j, m_j}$ is 
the spin-angular function~\cite{sakurai2011modern} 
satisfying
$\vecop{J}^2 \ket{l, j, m_j} = 
j(j+1) \hbar^2 \ket{l, j, m_j}$
and 
$\hat{J}_z \ket{l, j, m_j} = 
m_j \hbar \ket{l, j, m_j}$
($\vecop{J} = \vecop{L} + \vecop{S}$).

The spin-angular function can be explicitly written
in terms of the orbital angular momentum eigenstates
$\ket{l, m_l}$ satisfying
$\vecop{L}^2 \ket{l, m_l} = 
l(l+1) \hbar^2 \ket{l, m_l}$
and 
$\hat{L}_z \ket{l, m_l} = m_l \hbar \ket{l, m_l}$
($m_l=-l, \cdots ,l$)
and the spin angular momentum eigenstates
$\ket{\uparrow}$ and $\ket{\downarrow}$:
for $j = l + 1/2$,
\begin{equation}
\begin{aligned}
\ket{l, j, m_j} &= 
\sqrt{\frac{l + m_j + \frac{1}{2}}{2l+1}} \ket{l, m_j - \frac{1}{2}} \ket{\uparrow} \\ 
&+
\sqrt{\frac{l - m_j + \frac{1}{2}}{2l+1}} \ket{l, m_j + \frac{1}{2}} \ket{\downarrow},
\end{aligned}
\end{equation}
and for $j = \left|l - 1/2 \right|$,
\begin{equation}
\begin{aligned}
\ket{l, j, m_j}
&= 
\sqrt{\frac{l - m_j + \frac{1}{2}}{2l+1}} \ket{l, m_j - \frac{1}{2}} \ket{\uparrow} \\ 
&- 
\sqrt{\frac{l + m_j + \frac{1}{2}}{2l+1}} \ket{l, m_j + \frac{1}{2}} \ket{\downarrow}.
\end{aligned}
\end{equation}

Using the fact that 
the spin-angular function is 
an eigenstate of
$\vecop{L} \cdot \vecop{S}$,
$\hat{V}_{\mathrm{SL}}$
can be rewritten as the sum of 
the scalar-relativistic and 
SOC parts:~\cite{PhysRevB.21.2630, PhysRevB.26.4199}
\begin{equation}
\label{eqn:Vsl_SRSO}
\hat{V}_{\mathrm{SL}} =
\sum_{l=0}^{l_{\mathrm{max}}}
\ket{l} V_{l}^{\mathrm{SR}}(r)  \bra{l} +
\sum_{l=1}^{l_{\mathrm{max}}}
\ket{l} V_{l}^{\mathrm{SO}}(r) \, 
\vecop{L} \cdot \vecop{S} \bra{l}
,
\end{equation}
where $\ket{l} \bra{l}$ is 
the orbital angular momentum projector 
for a given $l$,
which is the sum of 
$\ket{l, m_l} \bra{l, m_l}$ over all $m_l$.
In Eq.~(\ref{eqn:Vsl_SRSO}), 
the radial potentials of 
the scalar-relativistic and SOC parts 
of $\hat{V}_{\mathrm{SL}}$
are given as
\begin{equation}
\label{eqn:Vsl_SR}
\begin{aligned}
&V_{l}^{\mathrm{SR}} (r) = 
\frac{l+1}{2l+1} V_{l, l + \frac{1}{2}}(r) + 
\frac{l}{2l+1} V_{l, \left|l - \frac{1}{2}\right|}(r)
\end{aligned}
\end{equation}
and
\begin{equation}
\label{eqn:Vsl_SO}
\begin{aligned}
&V_{l}^{\mathrm{SO}} (r) = 
\frac{2}{2l+1} 
\left[ V_{l, l + \frac{1}{2}}(r) - V_{l, \left|l - \frac{1}{2}\right|}(r) \right],
\end{aligned}
\end{equation}
respectively.
The scalar-relativistic potential,
$V_{l}^{\mathrm{SR}}(r)$,
includes 
the effects of 
the Darwin term and the mass-velocity term.~\cite{schiff1968quantum}
Hybertsen and Louie~\cite{PhysRevB.34.2920} considered 
the effects of the SOC potential,
$V_{l}^{\mathrm{SO}}(r)$,
on the spin-orbit splittings
in the band structure of semiconductors
within first-order perturbation theory and 
found good agreement with experiments.

For computational efficiency,
pseudopotentials of 
the fully-separable 
Kleinman-Bylander (KB) form~\cite{PhysRevLett.48.1425,PhysRevB.47.4238,PhysRevB.64.073106} 
are commonly used 
instead of those of the semi-local form.
The non-local part of a fully-relativistic KB pseudopotential can be written as
\begin{equation}
\label{eqn:Vkb}
\begin{aligned}
\hat{V}_{\mathrm{KB}} =
\sum_{l=0}^{l_{\mathrm{max}}} \sum_{j = \left| l - \frac{1}{2} \right|}^{l + \frac{1}{2}} \sum_{m_j = -j}^{j}
\ket{l,j,m_j} \ket{\beta_{l,j}} \bra{\beta_{l,j}} \bra{l,j,m_j},
\end{aligned}
\end{equation}
where
the radially non-local projector 
$\ket{\beta_{l,j}} \bra{\beta_{l,j}}$
is used instead of 
the radial potential $V_{l,j}(r)$.

Similarly to the case of the semi-local pseudopotential,
a fully-relativistic KB pseudopotential 
can be rewritten as the sum of
the scalar-relativistic and SOC parts:
\begin{equation}
\label{eqn:Vkb_SRSO}
\hat{V}_{\mathrm{KB}} = 
\sum_{l=0}^{l_{\mathrm{max}}} 
\ket{l} \hat{V}_{l}^{\mathrm{SR}} \bra{l} +
\sum_{l=1}^{l_{\mathrm{max}}} 
\ket{l} \hat{V}_{l}^{\mathrm{SO}} \, 
\vecop{L} \cdot \vecop{S} \bra{l},
\end{equation}
where the non-local potentials 
of the scalar-relativistic 
and SOC parts of $\hat{V}_{\mathrm{KB}}$ 
are defined as
\begin{equation}
\label{eqn:Vkb_SR}
\begin{aligned}
\hat{V}_{l}^{\mathrm{SR}} 
&=
\frac{l+1}{2l+1} 
\ket{\beta_{l, l + \frac{1}{2}}} \bra{\beta_{l, l + \frac{1}{2}}}\\
&+
\frac{l}{2l+1} 
\ket{\beta_{l, \left|l - \frac{1}{2}\right|}} \bra{\beta_{l, \left|l - \frac{1}{2}\right|}}
\end{aligned}
\end{equation}
and
\begin{equation}
\label{eqn:Vkb_SO}
\begin{aligned}
\hat{V}_{l}^{\mathrm{SO}} 
=
\frac{2}{2l+1}
&\Big( \ket{\beta_{l, l + \frac{1}{2}}} \bra{\beta_{l, l + \frac{1}{2}}}\\ 
&- \ket{\beta_{l, \left|l - \frac{1}{2} \right|}} \bra{\beta_{l, \left| l - \frac{1}{2} \right|}} \Big),
\end{aligned}
\end{equation}
respectively.

The fully-relativistic velocity operator 
that includes all the non-local effects
of the fully-relativistic KB pseudopotential 
is written as
\begin{equation}
\label{eqn:vfr}
\vecop{v}^{\fr} =
\vecop{v}^{\p}
+ \frac{i}{\hbar} \, 
\left[\hat{V}_{\mathrm{KB}}, \vecop{r} \right],
\end{equation}
where $\vecop{v}^{\p} ( = \vecop{p}/m)$
is introduced for notational convenience.
The commutator on the right-hand side of 
Eq.~(\ref{eqn:vfr})
can be separated into
scalar-relativistic and SOC parts.
By using the expressions 
in Eqs.~(\ref{eqn:Vkb_SRSO}), 
(\ref{eqn:Vkb_SR}), 
and (\ref{eqn:Vkb_SO}),
we define
the scalar-relativistic velocity operator 
that includes only 
the effects arising from 
the scalar-relativistic part of $\hat{V}_{\mathrm{KB}}$: 
\begin{equation}
\label{eqn:vsr}
\vecop{v}^{\sr} = 
\vecop{v}^{\p}
+ \frac{i}{\hbar} \sum_{l=0}^{l_{\mathrm{max}}} 
\left[ \ket{l} \hat{V}_{l}^{\mathrm{SR}} \bra{l} , \vecop{r} \right].
\end{equation}

Within this formalism,
the non-local effects of SOC
on the velocity operator
arise from the difference between
$\vecop{v}^{\fr}$ and $\vecop{v}^{\sr}$
which can be written as 
the sum of the commutators 
arising from $\hat{V}^{\so}_{l}$:
\begin{equation}
\label{eqn:vso}
\hat{\mathbf v}^{\so}
\equiv \vecop{v}^{\fr} - \vecop{v}^{\sr}
= \sum_{l=1}^{l_{\rm max}} 
\vecop{v}_{l}^{\so}
\end{equation}
where 
\begin{equation}
\label{eqn:vsol}
\vecop{v}_{l}^{\so} = 
i/\hbar 
\left[ \,\ket{l} \hat{V}_{l}^{\mathrm{SO}} \, \hat{\mathbf{L}} \cdot \hat{\mathbf{S}} \bra{l}, \hat{\mathbf{r}}\, \right].
\end{equation}

The optical matrix elements of our interest are
$\mel{f}
{\mathbf{e} \cdot \vecop{v}^{(\mathrm{FR/SR/\mathbf{p}})}}
{i}$,
where $\ket{i}$ and $\ket{f}$ are the initial and final
electronic states, respectively, 
and $\mathbf{e}$ is 
the polarization vector of the incident light.
We investigate the difference between 
the matrix elements of $\hat{\mathbf{v}}^{\fr}$
and $\hat{\mathbf{v}}^{\sr}$
in several systems having heavy elements such as W and Bi.
In the case of an isolated atom,
the initial and final states are 
the eigenstates of angular momentum operators
$\ket{n,l,j,m_j}$
where $n$ is the principal quantum number.
In periodic systems,
the initial and final states are 
the Bloch states in the valence band $\ket{v,\mathbf{k}}$ 
and those in the conduction band $\ket{c,\mathbf{k}}$,
respectively, 
where $\mathbf{k}$ is the crystal momentum,
and $v$ and $c$ are the indices of the valence and conduction bands, 
respectively.

The imaginary part of the dielectric function is calculated within the independent-particle random-phase approximation:
\begin{equation}
\label{eqn:eps}
\begin{aligned}
\mathrm{Im}\, 
&\varepsilon^{\mathrm{(\mathbf{p}/SR/FR)}} \left( \omega \right) \\
&= \frac{4 \pi}{\omega^2 \Omega N_{\mathbf{k}}}
\sum_{\mathbf{k}} \sum_{c, v}
\left| \mel{c,\mathbf{k}}{\mathbf{e} \cdot \hat{\mathbf v}^{\mathrm{(\mathbf{p}/SR/FR)}}}{v,\mathbf{k}} \right|^2 \\
&\times \delta \left( E_{c,\mathbf{k}} - E_{v,\mathbf{k}} - \hbar\omega \right),
\end{aligned}
\end{equation}
where $\omega$ is the frequency of the incident light, 
$\Omega$ is the volume of the unit cell, 
$N_{\mathbf{k}}$ is the number of $\mathbf{k}$ points in the Brillouin zone,
and $E_{v, \mathbf{k}}$ and $E_{c, \mathbf{k}}$ are 
the Kohn-Sham energy eigenvalues of $\ket{v,\mathbf{k}}$ 
and $\ket{c,\mathbf{k}}$, respectively.

From Eq.~(\ref{eqn:eps}), 
we can see that 
SOC affects the absoprtion spectra of materials
in two different ways:
($i$) SOC changes the Kohn-Sham energy eigenvalues, $E_{v,\mathbf{k}}$ and $E_{c,\mathbf{k}}$,
and eigenstates, $\ket{c, \mathbf{k}}$ and $\ket{v, \mathbf{k}}$, and 
($ii$) SOC gives an additional contribution,
$\vecop{v}^{\so}$ in Eq.~(\ref{eqn:vso}),
to the (fully-relativistic) velocity operator.
The focus of our work is on the second contribution.

In this work, 
we performed fully-relativistic DFT calculations within
the generalized gradient approximation~\cite{PhysRevLett.77.3865}
using the Quantum ESPRESSO package.~\cite{0953-8984-21-39-395502,giannozzi2017jpcm}
The optical matrix elements and the imaginary part of the dielectric function were calculated 
by using the Yambo code.~\cite{Marini20091392}
We modified the program so that 
we can construct both the scalar-relativistic and fully-relativistic velocity operators 
using the same set of fully-relativistic KB pseudopotentials.
All the fully-relativistic KB pseudopotentials used 
in this work were generated 
by using the ONCVPSP code.~\cite{PhysRevB.88.085117}
The generating parameters 
for the pseudopotentials were taken from 
the work of Schlipf and Gygi,~\cite{Schlipf201536} 
while slight modifications were made 
to get the fully-relativistic pseudopotential of Bi.

\begin{table}[]
\centering
\caption{
Pseudopotentials 
used in this work.  
The pseudization radii of 
the pseudo-wavefunctions 
with different orbital angular momenta, 
$r_s$, $r_p$, $r_d$, and $r_f$, 
are shown in units of the Bohr radius.
}
\label{table:pseudo}
\begin{tabular}{l l l c c c c}
\hline
&Core &Valence &$r_s$ &$r_p$ &$r_d$ &$r_f$\\
\hline
S    &[Ne] &$3s^2 3p^4$ &2.12 &1.51 &- &-\\
Se   &[Ar] $3d^{10}$ &$4s^2 4p^4$ &2.60 &2.71 &3.33 &-\\
Mo   &[Ar] $3d^{10}$ &$4s^2 4p^6 5s^2 4d^{4}$ &2.00 &2.50 &2.56 &-\\
W(1) &[Kr] $4d^{10}$ &$4f^{14} 5s^2 5p^6 6s^2 5d^4$ &2.12 &2.19 &1.88 &3.03\\
W(2) &[Kr] $4d^{10} 4f^{14}$ &$5s^2 5p^6 6s^2 5d^4$ &2.02 &1.93 &1.84 &2.73\\
Bi   &[Xe] $4f^{14}$ &$5d^{10} 6s^2 6p^3$ &3.19 &3.15 &3.00 &3.34\\
\hline
\end{tabular}
\end{table}

\section{Results and Discussion}

\subsection{Isolated atomic systems} \label{sec:atoms}

\begin{figure}
\begin{center}
\includegraphics[width=\columnwidth]{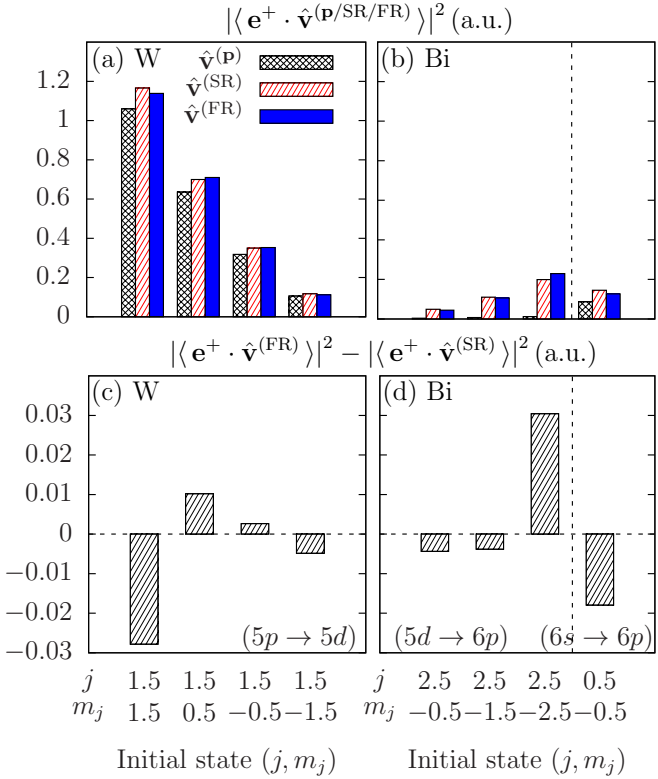}
\end{center}
\caption{
(a) and (b)
The squared optical matrix elements
of W and Bi atoms
obtained by using
the fully-relativistic velocity operator,
the scalar-relativistic velocity operator,
and the momentum operator
for the optical transitions
between the total angular momentum eigenstates.
Only the cases of 
the largest four squared matrix elements of 
the fully-relativistic velocity operator 
among $5p \rightarrow 5d$ transitions in a W atom 
and those among $5d \rightarrow 6p$ and $6s \rightarrow 6p$ transitions in a Bi atom
are shown.
(c) and (d) The difference between 
the squared optical matrix elements
obtained by using
the fully-relativistic velocity operator and
the scalar-relativistic velocity operator.
The incident light is left-circularly polarized.
The matrix elements are in Hartree.
}
\label{fig:atom_vmat}
\end{figure}

We study the effects of SOC 
on the optical matrix elements of isolated
W and Bi atoms
which are heavy elements and have an
electronic structure
strongly affected by SOC.
We calculated 
the squared optical matrix elements
of the form
$| \mel{f}{\mathbf{e}^{+} \cdot \vecop{v}^{(\mathrm{\mathbf{p}/SR/FR})}}{i} |^2$,
where the initial and final states are chosen to be 
the total angular momentum eigenstates,
$\ket{n,l,j,m_j}$,
and $\mathbf{e}^{+}$ is 
the polarization vector of 
the left-circularly polarized light propagating along 
the $z$ direction.
By comparing 
the squared matrix elements of 
$\vecop{v}^{\fr}$ and $\vecop{v}^{\sr}$ 
for a given pair of initial and final states,
we calculated 
the effects of SOC on 
the individual optical matrix element.

Figures~\ref{fig:atom_vmat}(a) and \ref{fig:atom_vmat}(b) 
show the squared matrix elements of 
$\vecop{v}^{\fr}$, $\vecop{v}^{\sr}$, and $\vecop{v}^{\p}$.
In the case of a W atom, 
the difference between 
the squared matrix elements of 
$\vecop{v}^{\sr}$ and 
$\vecop{v}^{\p}$  is not very large.
In the case of a Bi atom, 
however, 
the squared matrix elements of 
$\vecop{v}^{\sr}$ significantly differ from 
those of $\vecop{v}^{\p}$,
especially for 
the $5d \rightarrow 6p$ transitions.
It is known that
such non-local effects arising from
the scalar-relativistic part of $\hat{V}_{\mathrm{KB}}$
can be significant 
if there is a large difference between
the $l$-orbital components of
the scalar-relativistic potential, $\hat{V}^{\mathrm{SR}}_{l}$.~\cite{Marini20091392}
On the other hand, 
the difference between 
the squared matrix elements of $\hat{\mathbf{v}}^{\mathrm{(FR)}}$ and $\hat{\mathbf{v}}^{\mathrm{(SR)}}$ 
is relatively small
for all the optical transitions.

Figures~\ref{fig:atom_vmat}(c) and 
\ref{fig:atom_vmat}(d) 
show the difference between 
the squared matrix elements of 
$\vecop{v}^{\fr}$ and $\vecop{v}^{\sr}$.
In the case of a W atom,
the difference between the squared matrix elements
of $\vecop{v}^{\fr}$ and $\vecop{v}^{\sr}$
can be as large as 4.3~\% 
of the squared matrix elements of $\vecop{v}^{\fr}$.
This difference is even more significant for
a Bi atom and can reach 14~\%
(in the case of $6s \rightarrow 6p$ transitions).
Although
the effects of SOC on
the optical matrix elements
strongly depend upon the characters of
the initial and final states,
we find that
the non-local effects of SOC on the optical matrix elements
are not negligible in W and Bi atoms.

\begin{figure}
\begin{center}
\includegraphics[width=\columnwidth]{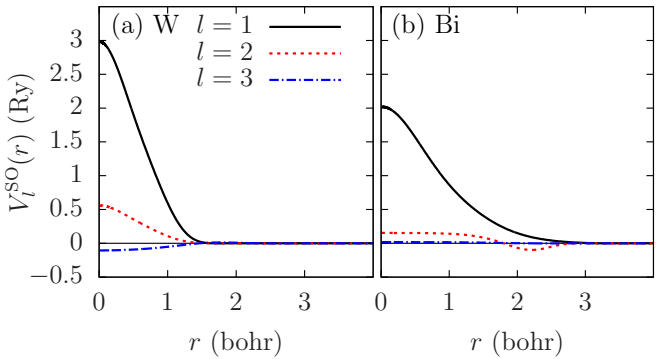}
\end{center}
\caption{
The spin-orbit coupling potential
[Eq.~(\ref{eqn:Vsl_SO})]
of the fully-relativistic pseudopotentials of
W [W(2) pseudopotential in Tab.~\ref{table:pseudo}]
and Bi atoms.
}
\label{fig:atom_pseudo}
\end{figure}

The relatively large effects of SOC
on the optical matrix elements
in the case of a Bi atom 
can be qualitatively understood 
by looking at
the SOC potentials
of the W and Bi pseudopotentials
in the semi-local form [Eq.~(\ref{eqn:Vsl_SO})].
Figure~\ref{fig:atom_pseudo} shows 
$V^{\mathrm{SO}}_{l}(r)$ 
of the W and Bi pseudopotentials used in our calculations
(their generating parameters are shown in Tab.~\ref{table:pseudo}).
Because $V_{l}^{\mathrm{SO}}(r)$ 
is defined as 
the difference between 
$V_{l,l+1/2}(r)$ and $V_{l,|l-1/2|}(r)$, 
$V_{l}^{\mathrm{SO}}(r)$ is localized within the pseudization radii of $V_{l,l+1/2}(r)$ and $V_{l,|l-1/2|}(r)$.
Because the pseudization radii of 
the W pseudopotential are smaller than 
those of the Bi pseudopotential 
(see Tab.~\ref{table:pseudo}),
$V_{l}^{\mathrm{SO}}(r)$ of the W pseudopotential 
are more localized than 
those of the Bi pseudopotential.
We note that for both atoms
the $p$-orbital part of
the SOC potential,
$V_{l=1}^{\mathrm{SO}}(r)$,
is much larger than 
the $d$- and $f$-orbital parts,
$V_{l=2}^{\mathrm{SO}}(r)$ and $V_{l=3}^{\mathrm{SO}}(r)$.

The matrix elements of 
$\hat{\mathbf{v}}^{\so}_{l}$
[Eq.~(\ref{eqn:vsol})]
can be explicitly written as
\begin{equation}
\label{eqn:vso_l}
\begin{aligned}
\mel{f}{ \vecop{v}^{\so}_{l}}{i} =
&\frac{i}{\hbar} \sum_{\sigma,\sigma'} 
\int d\mathbf{r} \, V_{l}^{\mathrm{SO}}(r) \, \times \\
&\psi_{f}^{*}(\mathbf{r},\sigma)
\left[ \vecop{L} \cdot \vecop{S}_{\sigma \sigma'} \ketbra{l}{l} , \mathbf{r} \right]
\psi_{i}(\mathbf{r},\sigma')
,
\end{aligned}
\end{equation}
where 
$\sigma$ and $\sigma'$ are the spin indices, 
$\psi_{i}(\mathbf{r}, \sigma)$ and 
$\psi_{f}(\mathbf{r} ,\sigma)$ are 
the $(\mathbf{r}, \sigma)$ component of 
$\ket{i}$ and $\ket{f}$.
Because Eq.~(\ref{eqn:vso_l}) contains 
the volume integration of $V_{l}^{\mathrm{SO}}(r)$,
not only the value of $V_{l}^{\mathrm{SO}}(r)$ 
near the core region $(r \sim 0)$ 
but also the spatial extent of $V_{l}^{\mathrm{SO}}(r)$ 
is an important factor that affects 
the magnitude of $\mel{f}{ \vecop{v}^{\so}_{l}}{i}$.

\begin{figure*}
\begin{center}
\includegraphics[width=2.0\columnwidth]{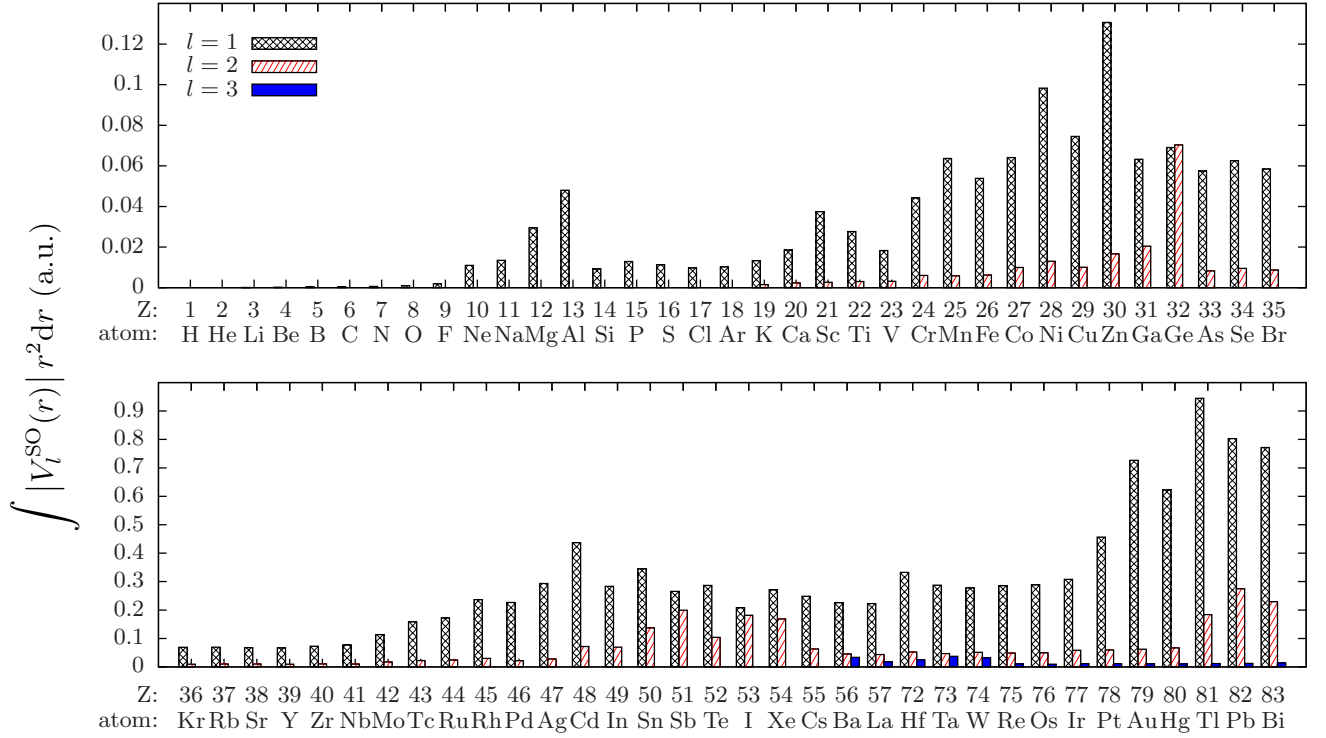}
\end{center}
\caption{
The volume integration of 
$|V_{l}^{\mathrm{SO}}(r)|$ [see Eq.~(\ref{eqn:Vsl_SO})] 
of the fully-relativistic pseudopotentials of
the atoms 
from H ($Z=1$) to Bi ($Z=83$) 
except those in the Lanthanide series.
}
\label{fig:atom_vso_integ}
\end{figure*}

To sketch 
the influence of $V^{\mathrm{SO}}_{l}(r)$
on the matrix elements of $\vecop{v}^{\so}_{l}$,
we evaluate 
the volume integration of $|V_{l}^{\mathrm{SO}}(r)|$ 
for the atoms in the periodic table 
from H ($Z=1$) to Bi ($Z=83$) 
except those in the Lanthanide series (Fig.~\ref{fig:atom_vso_integ}).
Roughly speaking, 
the volume integration of 
$|V_{l}^{\mathrm{SO}}(r)|$ 
increases with the atomic number $Z$, 
or with the atomic mass.
For example, the volume integration of 
$|{V}_{l=1}^{\mathrm{SO}}(r)|$ of 
the Bi ($Z=83$) pseudopotential is 0.78 Hartree, 
while the same quantity of the W ($Z=74$) pseudopotential is 0.3.
The result is consistent with 
our observation in Fig.~\ref{fig:atom_vmat} 
that 
the effects of SOC
on the optical matrix elements
increase with the atomic mass
(4.3~\% and 14~\% of 
the squared matrix elements of $\vecop{v}^{\fr}$
for W and Bi atoms, respectively).

For most atoms in Fig.~\ref{fig:atom_vso_integ},
we find that
the $p$-orbital component of the SOC potential,
$|V_{l=1}^{\mathrm{SO}}(r)|$, is the largest, 
the $d$-orbital component, $|V_{l=2}^{\mathrm{SO}}(r)|$, is the second largest,
and the $f$-orbital compoenent,
$|V_{l=3}^{\mathrm{SO}}(r)|$, is the smallest.
Therefore, 
it is natural to expect that 
the contribution of $\vecop{v}_{l=1}^{\so}$
to the optical matrix elements
is the most important one
and that
the contribution of $\vecop{v}^{\so}_{l}$
becomes smaller as $l$ increases.
To investigate 
the contribution of $\vecop{v}_{l}^{\so}$ 
to the squared matrix elements of $\vecop{v}^{\fr}$, 
we calculated 
the squared matrix elements of 
$\vecop{v}^{\fr}$ and 
$\vecop{v}^{\fr} - \vecop{v}^{\so}_{l}$.

\begin{figure}
\begin{center}
\includegraphics[width=\columnwidth]{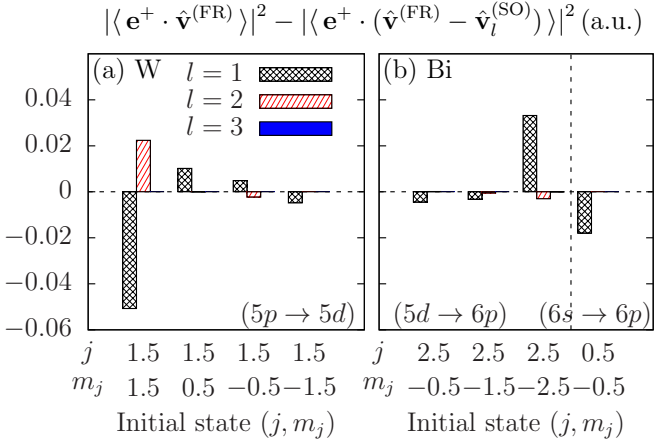}
\end{center}
\caption{
The difference between 
the squared matrix elements
of W and Bi atoms
obtained by using
the fully-relativistic velocity operator and
$\hat{\mathbf{v}}^{\mathrm{(FR)}} - \hat{\mathbf{v}}^{\mathrm{(SO)}}_{l}$ 
[see Eqs.~(\ref{eqn:vfr}) and (\ref{eqn:vsol})].
The initial and final states are 
the same as those in Fig.~\ref{fig:atom_vmat}.
}
\label{fig:atom_vmat_ldep}
\end{figure}

Figure~\ref{fig:atom_vmat_ldep} shows 
the difference between 
the squared matrix elements of 
$\vecop{v}^{\fr}$ and $\vecop{v}^{\fr} - \vecop{v}^{\so}_{l}$ 
in the cases of W and Bi atoms.
In both cases,
the contribution of  $\vecop{v}^{\so}_{l=1}$
to the squared matrix elements of $\vecop{v}^{\fr}$
is the largest
among the contributions of $\vecop{v}^{\so}_{l}$.
The contribution of $\vecop{v}^{\so}_{l=2}$
is the second largest
and that of $\vecop{v}^{\so}_{l=3}$
is the smallest and negligible.
In the case of a W atom, 
there is a case where
the contribution of $\vecop{v}^{\so}_{l=2}$ 
is almost half of 
the contribution of $\vecop{v}^{\so}_{l=1}$.
In the case of a Bi atom, 
the contributions of 
$\vecop{v}^{\so}_{l=2}$ and $\vecop{v}^{\so}_{l=3}$
are very small.

Because ${\vecop{v}^{\so}_{l}}$ contains
the orbital angular momentum projector, $\ket{l} \bra{l}$,
the matrix elements of ${\vecop{v}^{\so}_{l}}$
are finite only
if the initial state or the final state
has $l$-orbital angular momentum character.
In addition,
according to the optical selection rule,
the matrix elements $\mel {f} {\vecop{v}^{\so}_{l}} {i}$
are finite 
if 
the difference between the $l$'s of
$\ket{i}$ and $\ket{f}$ is $\pm 1$.
Therefore, 
the matrix elements of $\vecop{v}^{\so}_{l=3}$
are finite only for
the transitions between 
the states 
that have 
$d$- and $f$-orbital angular momentum characters.
In our calculations, 
the contribution of ${\vecop{v}^{\so}_{l=3}}$
to the squared matrix elements of $\vecop{v}^{\fr}$
is usually very small for such transitions.

It is known that the effects of SOC on
the energy levels and wavefunctions of atomic systems
are the largest for $p$-orbitals and
become smaller for $d$- and $f$-orbitals.~\cite{burke1967effect}
In fact, if we recall the fine structure of a hydrogen atom,
we easily see that
the energy splittings induced by SOC
show the same $l$-dependent behavior
($p>d>f$).~\cite{griffiths2005introduction}\textsuperscript{,}\footnote{see, for example, Fig.~6.9 of Ref.~\onlinecite{griffiths2005introduction}}
Our results are in line with these
relativistic effects on atomic systems.

\subsection{Monolayer transition metal dichalcogenides}

\begin{figure}
\begin{center}
\includegraphics[width=\columnwidth]{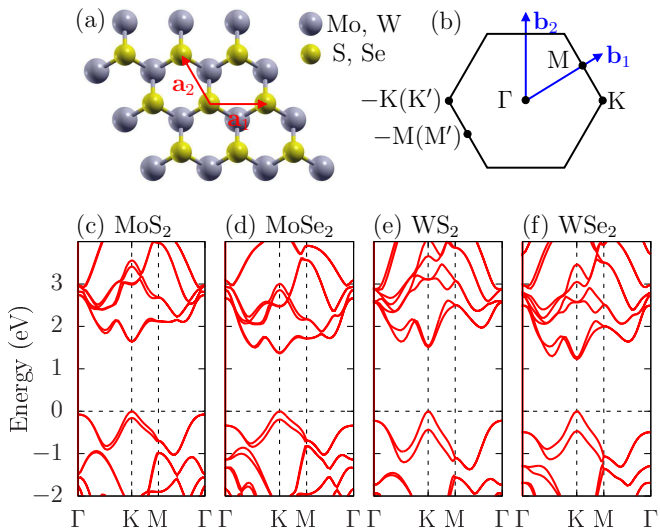}
\end{center}
\caption{
(a) and (b) The crystal structure and the Brillouin zone 
of a monoayer of 
2H-type semiconducting 
transition metal dichalcogenides, respectively.
(c)-(f) The electronic band structure of 
a monolayer of four 2H-type transition metal dichalcogenides.
}
\label{fig:tmdc_struct}
\end{figure}

We calculate the optical matrix elements 
and absorption spectra of 
a monolayer of four 2H-type semiconducting TMDCs 
(MoS$_2$, MoSe$_2$, WS$_2$, and WSe$_2$).
Figure~\ref{fig:tmdc_struct}(a) 
shows the structure of the two-dimensional crystal.
The TMDC monolayer of 2H-type consists of 
a transition metal layer (Mo or W) 
which is sandwiched by two chalcogen layers (S or Se).
Figure~\ref{fig:tmdc_struct}(b) shows 
the two-dimensional Brillouin zone.

In our DFT calculations 
using fully-relativistic pseudopotentials, 
we set the kinetic energy cutoff of the plane-wave basis 
to 80~Ry and 
sample the Brilluoin zone with a 
$12 \times 12 \times 1$ Monkhorst-Pack grid.~\cite{PhysRevB.13.5188}
In calculations of absorption spectra, 
we use a denser $120 \times 120 \times 1$ Monkhorst-Pack grid for $\mathbf{k}$-point summations.

Figures~\ref{fig:tmdc_struct}(c)-(f) show 
the electronic band structure of TMDC monolayers.
At the valence-band maximum at K, 
we see energy splittings 
between the two highest bands in valence 
thanks to the large SOC of 4d and 5d transition metals.
The magnitude of the SOC-induced energy splittings 
varies from 153 to 468~meV 
(larger for the TMDCs having W atoms).
All the results of our calculations are consistent 
with previous theoretical studies.~\cite{PhysRevB.84.153402, PhysRevB.90.245411}

\begin{figure*}
\begin{center}
\includegraphics[width=2.0\columnwidth]{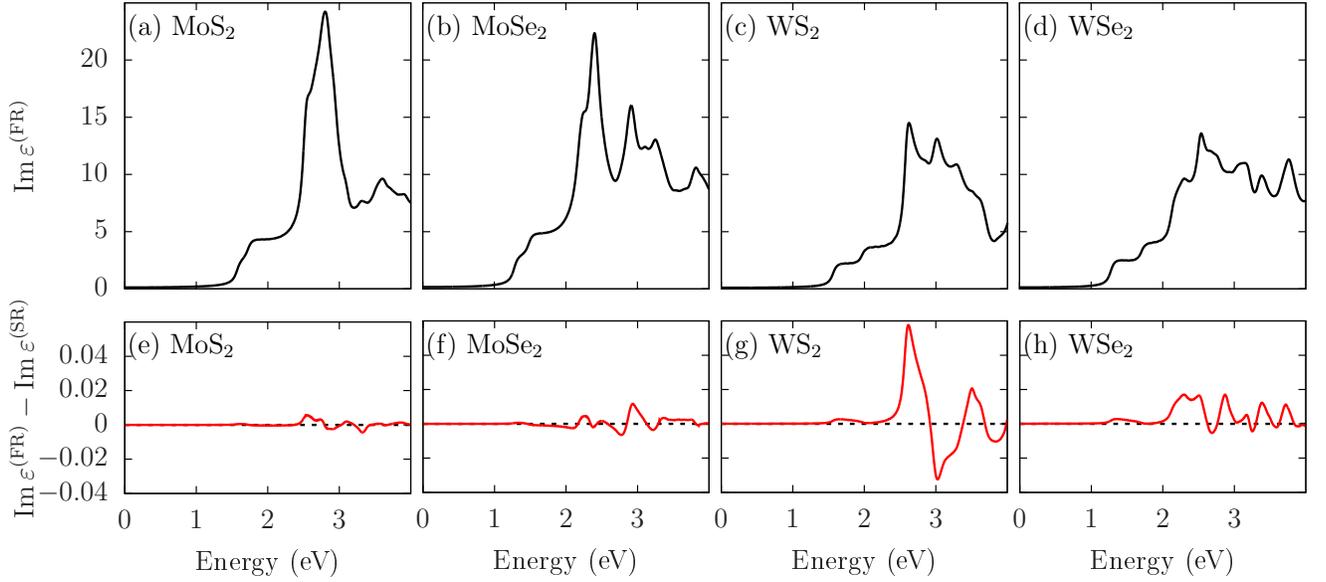}
\end{center}
\caption{
(a)-(d) The imaginary part of the dielectric function
of monolayer transition metal dichalcogenides 
obtained by using the fully-relativistic velocity operator.
(e)-(h) The difference 
between the imaginary part of 
the dielectric functions of monolayer transition metal dichalcogenides 
obtained by using 
the fully-relativistic velocity operator 
and the scalar-relativistic velocity operator.
}
\label{fig:tmdc_eps}
\end{figure*}

To investigate
the effects of SOC
on the absorption spectra of monolayer TMDCs,
we calculated the imaginary part
of the independent-particle 
dielectric function
using $\vecop{v}^{\fr}$,
$\vecop{v}^{\sr}$, and $\vecop{v}^{\p}$:
$\mathrm{Im}\,\varepsilon^{\fr}(\omega)$,
$\mathrm{Im}\,\varepsilon^{\sr}(\omega)$,
and $\mathrm{Im}\,\varepsilon^{\p}(\omega)$.
Figures~\ref{fig:tmdc_eps}(a)-(d)
show $\mathrm{Im}\,\varepsilon^{\fr}(\omega)$
of monolayer TMDCs.
The onset energies of 
$\mathrm{Im}\,\varepsilon^{\fr}(\omega)$ 
correspond to 
the band gaps of monolayer TMDCs.
In the low energy regime 
($\hbar\omega < 2$~eV), 
we see step-function-like behaviors 
of $\mathrm{Im}\,\varepsilon^{\fr}(\omega)$ 
which mainly result from 
the optical transitions 
between the band-edge states at K and K$'$, i.e.
the two highest-energy valence and 
the two lowest-energy conduction bands.
These band-edge states mostly consist of
the $d$ orbitals of transition metal atoms and
the $p$ orbitals of chalcogen atoms.
In the high energy regime ($\hbar\omega > 2$~eV), 
additional sharp features arise
as the states other than the band-edge states 
make contributions 
to $\mathrm{Im}\,\varepsilon^{\fr}(\omega)$.

Figures~\ref{fig:tmdc_eps}(e)-(h) 
show the difference 
between $\mathrm{Im}\,\varepsilon^{\fr}(\omega)$ 
and $\mathrm{Im}\,\varepsilon^{\sr}(\omega)$.
In the low energy regime 
($\hbar\omega < 2$~eV),
the difference between 
$\mathrm{Im}\,\varepsilon^{\fr}(\omega)$ 
and $\mathrm{Im}\,\varepsilon^{\sr}(\omega)$
is three orders of magnitudes smaller than 
$\mathrm{Im}\,\varepsilon^{\fr}(\omega)$ itself,
which indicates that 
the non-local effects of SOC 
on $\mathrm{Im}\,\varepsilon^{\fr}(\omega)$ 
are negligibly small 
in this range of energy. 
The effects of SOC become larger 
at higher energies ($\hbar\omega > 2$~eV).
However, 
the difference between 
$\mathrm{Im}\,\varepsilon^{\fr}(\omega)$ and
$\mathrm{Im}\,\varepsilon^{\sr}(\omega)$ 
remains smaller 
than 1~\% of 
$\mathrm{Im}\,\varepsilon^{\fr}(\omega)$.

In general, 
the non-local part of 
a pseudopotential
strongly depends on 
its generating parameters such as 
valence (and core) configurations, 
pseudization radii, 
and the local part of the pseudopotential.
Therefore, 
the effects of 
the commutator on 
the optical matrix elements and absorption spectra
of a material
(including the contributions from 
the scalar-relativistic and SOC parts of the pseudopotential)
can change significantly
by the non-local character of 
the pseudopotential used in the calculations.

\begin{figure}
\begin{center}
\includegraphics[width=\columnwidth]{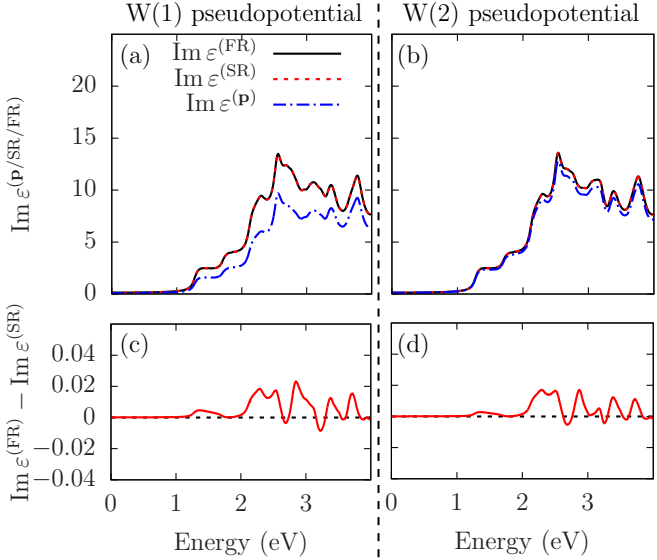}
\end{center}
\caption{
(a) and (b) 
The imaginary part of 
the dielectric function of monolayer WSe$_2$ obtained by using two different pseudopotentials of W, which were generated from two different valence configurations (see Tab.~\ref{table:pseudo} for details). 
$\textrm{Im}\,\varepsilon^{\p}(\omega)$,
$\textrm{Im}\,\varepsilon^{\sr}(\omega)$, 
and
$\textrm{Im}\,\varepsilon^{\fr}(\omega)$ are 
the imaginary part of the dielectric function 
obtained by using
the momentum operator, 
the scalar-relativistic velocity operator, 
and the fully-relativistic velocity operator, 
respectively.
(c) and (d) 
The difference between 
the imaginary part of 
the dielectric function 
obtained by using 
the fully-relativistic and 
the scalar-relativistic velocity operators.
}
\label{fig:wse2_pseudo_eps}
\end{figure}

In Fig.~\ref{fig:wse2_pseudo_eps},
we plot  
the imaginary part of the dielectric function
of monolayer WSe$_2$ 
obtained by using 
two different fully-relativistic KB pseudopotentials of W 
(for comparison, we fixed the pseudopotential of Se).
We checked that 
the two different pseudopotentials of W 
yield almost the same band structure 
within the energy range of our interest 
($\hbar\omega =$ 0--4~eV).
The absorption spectra 
in Figs.~\ref{fig:wse2_pseudo_eps}(a) and \ref{fig:wse2_pseudo_eps}(c) 
were obtained by using a W pseudopotential 
that includes $4f$ electrons in the valence 
[see W(1) in Tab.~\ref{table:pseudo}],
while those in Fig.~\ref{fig:wse2_pseudo_eps}(b) and \ref{fig:wse2_pseudo_eps}(d) 
were obtained by using a W pseudopotential 
that includes $4f$ electrons in the core 
[see W(2) in Tab.~\ref{table:pseudo}].

By comparing 
$\mathrm{Im}\,\varepsilon^{\p}(\omega)$
in Figs.~\ref{fig:wse2_pseudo_eps}(a) and \ref{fig:wse2_pseudo_eps}(b), 
we find that the absorption spectrum
strongly depends on the pseudopotential of W, 
if we neglect all the effects arising from 
the non-local part of the pseudopotential and 
use $\vecop{v}^{\p}$ as the velocity operator.
In the case of the W(1) pseudopotential,
$\mathrm{Im}\,\varepsilon^{(\mathbf{p})}(\omega)$ 
is much smaller than 
$\mathrm{Im}\,\varepsilon^{\sr}(\omega)$.
The result shows that the commutator arising from 
the scalar-relativistic part of 
the W(1) pseudopotential 
affects strongly the absorption spectrum. 
The large difference between
$\mathrm{Im}\,\varepsilon^{\sr} (\omega)$ and
$\mathrm{Im}\,\varepsilon^{\p}(\omega)$ 
is attributed to 
the presence of $4f$ electrons in the valence 
which makes the W(1) pseudopotential strongly non-local.
On the other hand, 
in the case of the W(2) pseudopotential,
$\mathrm{Im}\,\varepsilon^{\p}(\omega)$ 
is quite similar to $\mathrm{Im}\,\varepsilon^{\sr}(\omega)$.
The effects of the commutator arising 
from the scalar-relativistic part of 
the pseudopotential
are much smaller for 
the W(2) pseudopotential 
where $4f$ electrons are treated as core electrons.
We note that in both cases, 
$\mathrm{Im}\,\varepsilon^{\sr}(\omega)$ 
is almost identical to 
$\mathrm{Im}\,\varepsilon^{\fr}(\omega)$.

Figures~\ref{fig:wse2_pseudo_eps}(c) 
and \ref{fig:wse2_pseudo_eps}(d) 
show the difference between 
$\mathrm{Im}\,\varepsilon^{\fr}(\omega)$ 
and $\mathrm{Im}\,\varepsilon^{\sr}(\omega)$.
We find that 
the contributions of SOC
to $\mathrm{Im}\,\varepsilon^{\fr}(\omega)$
for the two different pseudopotentials of W
are very similar to each other
in the whole energy range.
The result shows that 
the effects of 
the commutator arising from SOC 
on the absorption spectra 
do not depend much on 
the pseudopotential.

\begin{figure*}
\begin{center}
\includegraphics[width=2.0\columnwidth]{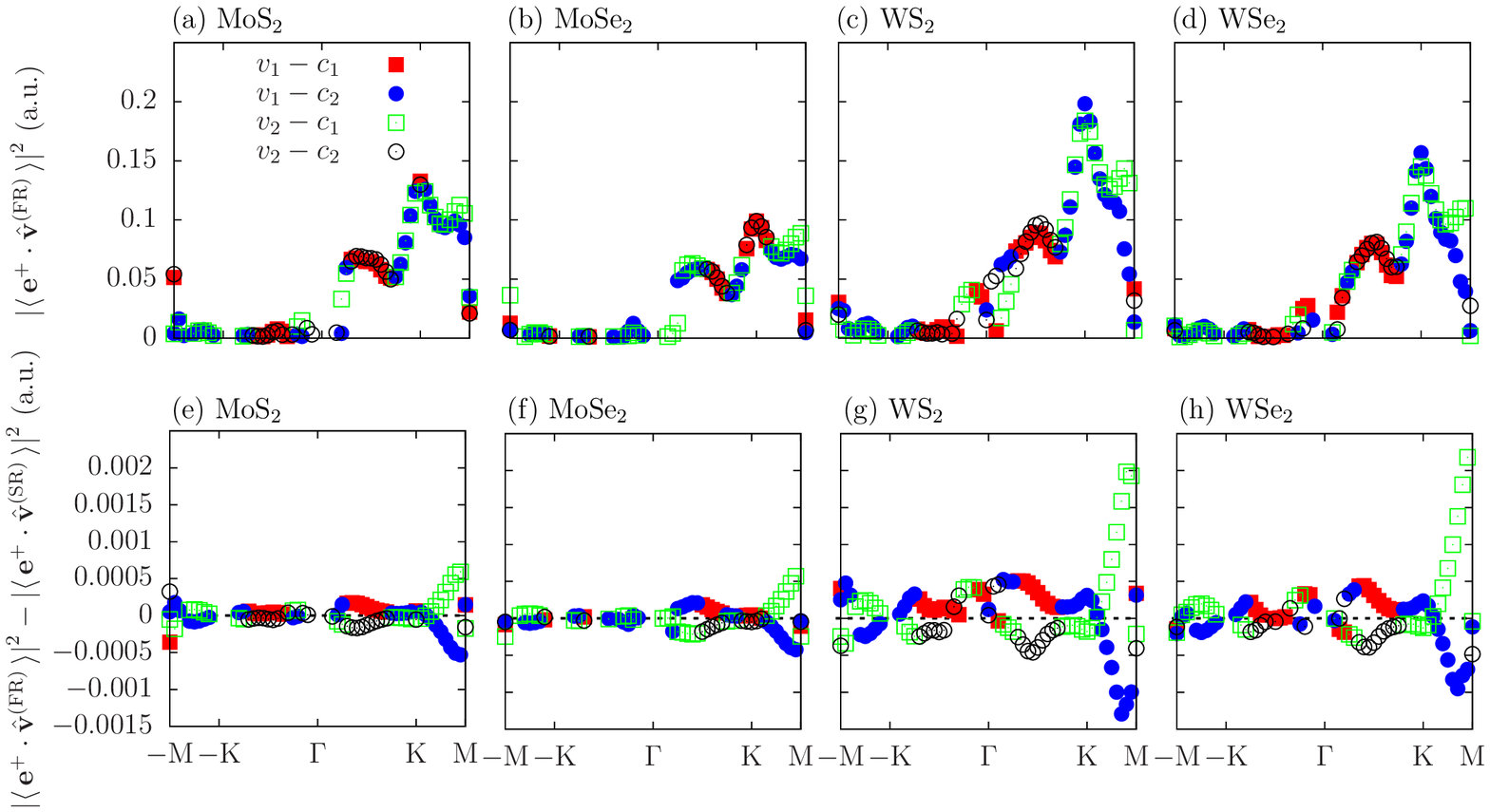}
\end{center}
\caption{
(a)-(d) The squared optical matrix elements 
obtained by using the fully-relativistic velocity operator
of monolayer transition metal dichalcogenides
for the optical transitions involving 
the highest-energy and second-highest-energy states
in the valence band,
$v_1$ and $v_2$, respectively,
and the lowest-energy and second-lowest-energy states
in the conduction band,
$c_1$ and $c_2$, respectively. 
The squared optical matrix elements were calculated 
along the path in the momentum space,
$-\mathrm{M} \rightarrow -\mathrm{K} \rightarrow \Gamma \rightarrow \mathrm{K} \rightarrow \mathrm{M}$.  
(e)-(h) The difference between 
the squared optical matrix elements 
obtained by using the fully-relativistic velocity operator 
and those obtained 
by using the scalar-relativistic velocity operator.  
In all cases, the incident light is left-circularly polarized.
}
\label{fig:tmdc_vmat_band_edge}
\end{figure*}

It is possible that even if
the non-local effects of SOC 
on the individual optical matrix element are large,
the effects 
on the absorption spectrum
are small
as we sum over the contributions from 
many optical matrix elements 
with different momenta and band indices.
To check this possibility, 
we calculated the squared matrix elements 
$|\mel{c_{i},\mathbf{k}}
{\mathbf{e}^{+} \cdot \vecop{v}^{\mathrm{(SR/FR)}}}
{v_{j},\mathbf{k}}|^2$, 
where $i$ and $j$ are 1 or 2, 
$v_1$ and $v_2$ are 
the band indices of the highest-energy and
second-highest-energy states 
in the valence band, respectively,
$c_1$ and $c_2$ are 
the band indices of the lowest-energy and
second-lowest-energy states 
in the conduction band, respectively,
and $\mathbf{k}$ is on the path 
$-\mathrm{M} \rightarrow -\mathrm{K} \rightarrow \Gamma \rightarrow \mathrm{K} \rightarrow \mathrm{M}$ [Fig.~\ref{fig:tmdc_struct}(b)].

Figures~\ref{fig:tmdc_vmat_band_edge}(a)-(d) show 
the squared matrix elements of $\vecop{v}^{\fr}$.
Here,
we see that 
the squared matrix elements of $\vecop{v}^{\fr}$ 
near K are larger in magnitude
than those near $-\mathrm{K}$.
Because we assumed
the incident light to be left-circularly polarized, 
the result can be explained by
the valley-selective circular dichroism
of monolayer TMDCs.~\cite{Cao2012}

Figures~\ref{fig:tmdc_vmat_band_edge}(e)-(h) show 
the difference between 
the squared matrix elements of
$\vecop{v}^{\fr}$ and $\vecop{v}^{\sr}$.
Although
the contribution of the commutator arising from SOC 
to the squared matrix elements of $\vecop{v}^{\fr}$
becomes larger in the case of having
heavier transition metal atoms (WS$_2$ and WSe$_2$),
even in those cases
the contribution from SOC
remains smaller than 
1~\% of the squared matrix elements of $\vecop{v}^{\fr}$.
If we compare this result with 
the previous result of an isolated W atom,
the influence of 
the commutator arising from SOC 
on the optical matrix elements
is much suppressed:
In the case of a W atom, 
the effects of SOC on 
the squared optical matrix elements
can be as large as 
4.3~\% of the squared matrix elements of 
$\vecop{v}^{\fr}$ [Figs.~\ref{fig:atom_vmat}(a) and \ref{fig:atom_vmat}(c)].

\begin{figure*}
\begin{center}
\includegraphics[width=1.7\columnwidth]{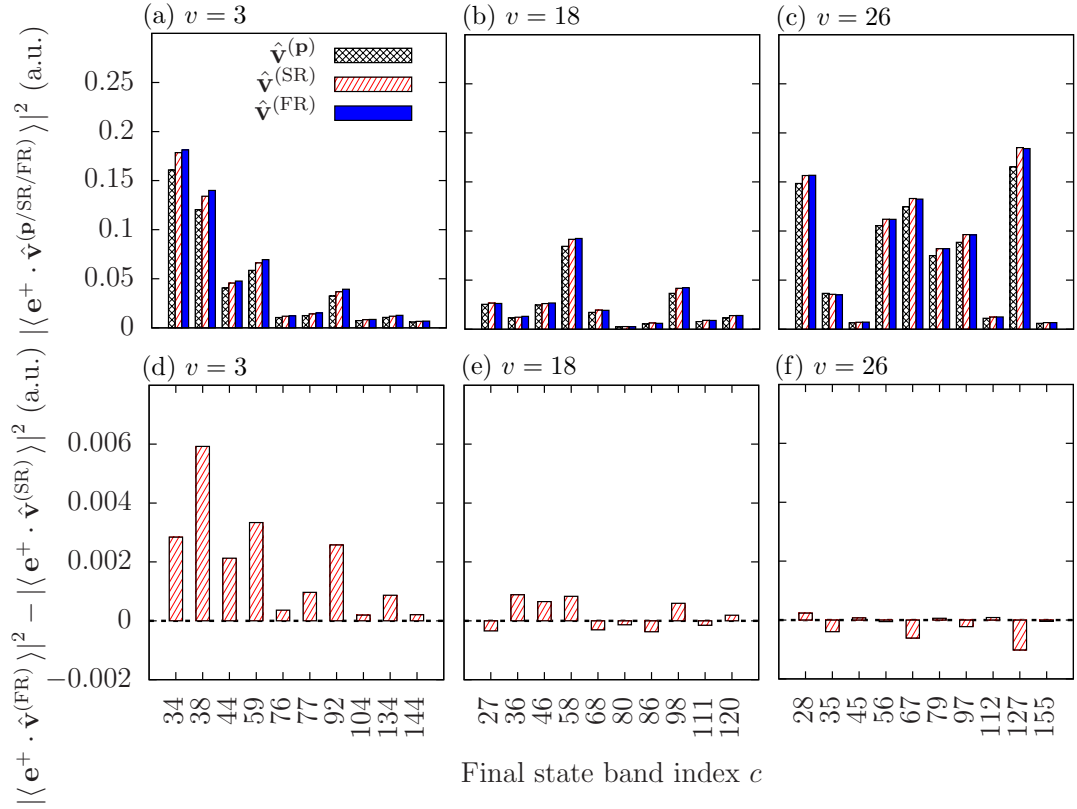}
\end{center}
\caption{
(a)-(c) The squared optical matrix elements 
of monolayer WSe$_2$
obtained by using 
the fully-relativistic velocity operator, 
the scalar-relativistic operator,
and the momentum operator 
for the optical transitions at K in the momentum space.  
(d)-(f) The difference between 
the squared optical matrix elements obtained 
by using the fully-relativistic velocity operator and 
the scalar-relativistic operator.  
In all cases, 
left-circularly polarized light was considered.
}
\label{fig:wse2_vmat}
\end{figure*}

Next, we further investigate the dependence of 
the squared matrix elements of
$\vecop{v}^{\fr}$, $\vecop{v}^{\sr}$, and $\vecop{v}^{\p}$
on the initial and final states
in the case of monolayer WSe$_2$.
We calculated the squared matrix elements
at K,
$|\mel{c,\mathrm{K}} 
{\mathbf{e}^{+} \cdot \vecop{v}^{\mathrm{(\mathbf{p}/SR/FR)}}}
{v,\mathrm{K}}|^2$,
where $v$ and $c$ are 
the band indices of the initial and final states, 
respectively.
The band indices are in increasing order of energy
(the state at the valence band maximum is $v=26$).
We consider three initial states
having different orbital characters:
($i$) $\ket{3, \mathrm{K}}$ 
which mostly consists of 
the $5p$ orbitals of W atoms,
($ii$) $\ket{18, \mathrm{K}}$ 
which consists of 
the $5p$ and $4d$ orbitals of W atoms and 
the $3p$ orbitals of Se atoms, and
($iii$) $\ket{26, \mathrm{K}}$ 
which 
consists of the $4d$ orbitals of W atoms and 
the $3p$ orbitals of Se atoms.
For each initial state,
we consider all the final states satisfying
$E_{c,\mathrm{K}} - E_{v,\mathrm{K}} < 1$~Ry.

Figures~\ref{fig:wse2_vmat}(a)-(c) show 
the squared matrix elements of 
$\vecop{v}^{\fr}$, $\vecop{v}^{\sr}$, 
and $\vecop{v}^{\p}$
and Figs.~\ref{fig:wse2_vmat}(d)-(f) show 
the difference between 
the squared matrix elements of $\vecop{v}^{\fr}$ 
and $\vecop{v}^{\sr}$.
By comparing the results of 
Figs.~\ref{fig:wse2_vmat}(d), \ref{fig:wse2_vmat}(e),
and \ref{fig:wse2_vmat}(f), 
we find that 
the effects of SOC on 
the optical matrix elements
are larger for
the optical transitions
whose initial state is 
more localized at W atoms.
In the case of $\ket{3, \mathrm{K}}$,
the effects of SOC on 
the squared matrix elements of $\vecop{v}^{\fr}$ 
can be as large as 6.8~\%,
while in the case of $\ket{26, \mathrm{K}}$,
the effects
are much smaller, 
less than 1.1~\%. 
The result of $\ket{18, \mathrm{K}}$
falls somewhere between the results of 
$\ket{3, \mathrm{K}}$ and $\ket{26, \mathrm{K}}$.

\begin{figure*}
\begin{center}
\includegraphics[width=1.5\columnwidth]{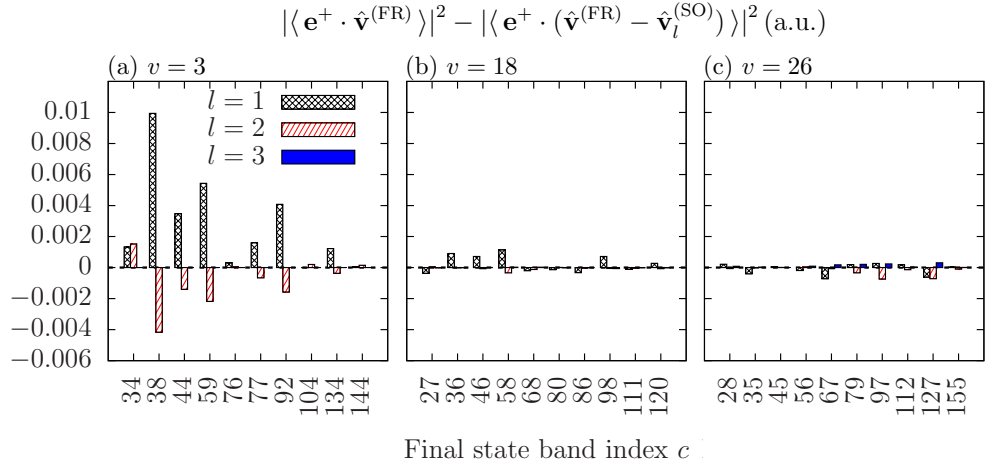}
\end{center}
\caption{
(a)-(c) The difference between 
the squared matrix elements of monolayer WSe$_2$ 
obtained by using 
the fully-relativistic velocity operator and 
$\hat{\mathbf{v}}^{\mathrm{(FR)}} - \hat{\mathbf{v}}^{\mathrm{(SO)}}_{l}$ [see Eqs.~(\ref{eqn:vfr}) and (\ref{eqn:vso})] for the optical transitions at K in the momentum space.
In all cases, 
left-circularly polarized light was considered.
}
\label{fig:wse2_vmat_ldep}
\end{figure*}

Figure~\ref{fig:wse2_vmat_ldep} show 
the differences between 
the squared matrix elements of 
$\vecop{v}^{\fr}$ and 
$\vecop{v}^{\fr} - \vecop{v}^{\so}_{l}$
for the optical transitions having 
$\ket{3, \mathrm{K}}$, 
$\ket{18, \mathrm{K}}$, 
and $\ket{26, \mathrm{K}}$ 
as the initial states.
In the case of $\ket{3,\mathrm{K}}$,
we find that 
among $\vecop{v}^{\so}_{l}$'s 
$\vecop{v}^{\so}_{l=1}$
gives the largest contribution to 
the squared matrix elements of $\vecop{v}^{\fr}$.
The $d$-orbital part $\vecop{v}^{\so}_{l=2}$ gives 
the second largest contribution and 
the contribution from 
the $f$-orbital part $\vecop{v}^{\so}_{l=3}$ is negligible.
This result is similar to
the result of an isolated W atom [Fig.~\ref{fig:atom_vmat_ldep}(a)].

Figures~\ref{fig:wse2_vmat_ldep}(b) and 
\ref{fig:wse2_vmat_ldep}(c) show 
that $\vecop{v}^{\so}_{l=1}$ 
is relatively less important 
in the cases of $\ket{18,\mathrm{K}}$ and $\ket{26,\mathrm{K}}$
than in the case of $\ket{3,\mathrm{K}}$.
The result can be qualitatively understood by looking at 
the W $p$- and $d$-orbital characters of the initial states.
As we move 
from $\ket{3,\mathrm{K}}$ to 
$\ket{18,\mathrm{K}}$ and 
$\ket{26,\mathrm{K}}$, 
the proportion of
the W $5p$-orbital component in the initial state
decreases while 
that of the W $4d$-orbital component increases.
In the case of $\ket{26,\mathrm{K}}$,
because the initial state mostly consists of 
the W $4d$ orbitals,
the matrix elements of $\vecop{v}^{\so}_{l=1}$
are finite
only for the final states having W $6p$-orbital character
($\Delta l = \pm 1$).
Such final states are 
much more delocalized
than the initial state
and the matrix elements of $\vecop{v}^{\so}_{l=1}$ are small.

\subsection{Bi$_2$Se$_3$ and Bi$_2$Te$_3$}

\begin{figure}
\begin{center}
\includegraphics[width=\columnwidth]{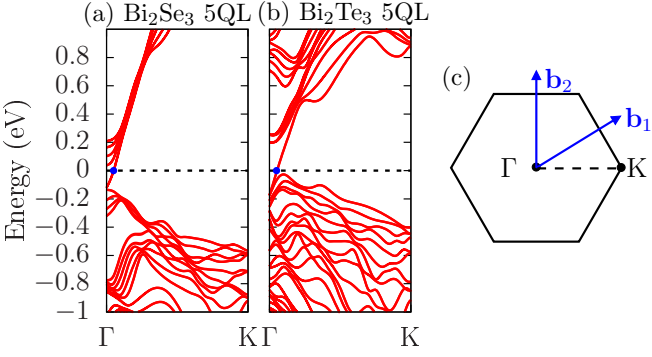}
\end{center}
\caption{
(a) and (b) The electronic band structure of 
5-quintuple-layer slabs of Bi$_2$Se$_3$ and Bi$_2$Te$_3$.
(c) The Brillouin zone of 
5-quintuple-layer slabs of Bi$_2$Se$_3$ and Bi$_2$Te$_3$.
}
\label{fig:ti_band_struct}
\end{figure}

We investigate the effects of SOC 
on the optical matrix elements of 
5-quintuple-layer slabs of 
Bi$_2$Se$_3$ and Bi$_2$Te$_3$.
Here, we focus on the optical transitions 
whose initial states are the topological surface states.
In DFT calculations, 
we slightly broke the inversion symmetry 
to induce small energy splittings 
between the surface states 
localized at the top and bottom sides of 
Bi$_2$Se$_3$ and Bi$_2$Te$_3$ slabs.
In this way, 
we can obtain the surface states $\ket{v, \mathbf{k}}$
($v$ is the band index and 
$v$ = 241 and 391 for Bi$_2$Se$_3$ and Bi$_2$Te$_3$,
respectively)
which are localized well on each surface of the slabs.
In the calculation of the optical matrix elements, 
we chose the surface state 
with momentum $\mathbf{k}' = 0.05\Gamma\mathrm{K}$
as our initial state
[blue dots in Figs.~\ref{fig:ti_band_struct}(a) and ~\ref{fig:ti_band_struct}(b)].
Also here, we consider all the final states
that satisfy
$E_{c, \mathbf{k}'} - E_{v, \mathbf{k}'} < 1$~Ry.
We set the kinetic energy cutoff of 
the plane-wave basis to
80~Ry and use a $6 \times 6 \times 1$ 
Monkhorst-Pack grid for $\mathbf{k}$-point sampling.

\begin{figure}
\begin{center}
\includegraphics[width=\columnwidth]{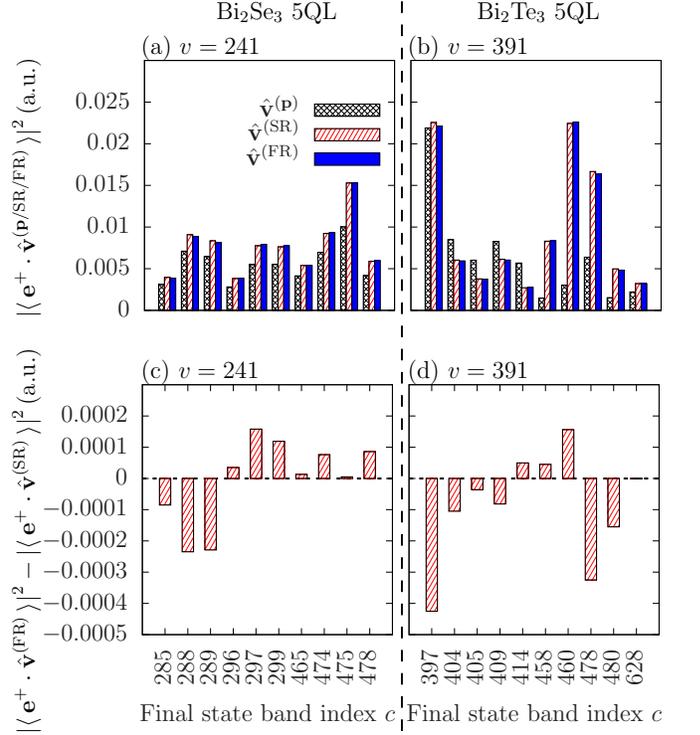}
\end{center}
\caption{
(a) and (b) 
The squared optical matrix elements 
of 5-quintuple-layer slabs of 
Bi$_2$Se$_3$ and Bi$_2$Te$_3$ 
obtained by using 
the fully-relativistic velocity operator, 
the scalar-relativistic operator, 
and the momentum operator 
for the optical transitions 
having the topological surface state with momentum $\mathbf{k}=0.05 \Gamma$K as the initial state.  
(c) and (d) 
The difference between 
the squared optical matrix elements 
obtained by using 
the fully-relativistic velocity operator and 
the scalar-relativistic operator.  
In all cases, 
left-circularly polarized light was considered.
}
\label{fig:ti_vmat}
\end{figure}

Figures~\ref{fig:ti_vmat}(a) and (b) show 
the squared matrix elements of 
$\vecop{v}^{\fr}$, $\vecop{v}^{\sr}$, 
and $\vecop{v}^{(\mathbf{p})}$.
We find that
the difference between
the squared matrix elements of 
$\vecop{v}^{\fr}$ and $\vecop{v}^{\sr}$
is very small,
while the difference between
the squared matrix elements of 
$\vecop{v}^{\sr}$ and $\vecop{v}^{\p}$
is large in some cases of Bi$_2$Te$_3$.
Figures~\ref{fig:ti_vmat}(c) and (d) show 
the differences between 
the squared matrix elements of 
$\vecop{v}^{\fr}$ and $\vecop{v}^{\sr}$
in a different scale.
As in the case of
the transitions from the valence-band maximum of monolayer WSe$_2$ [Figs.~\ref{fig:wse2_vmat}(c) and (f)], 
the effects of SOC on 
the optical matrix elements of 
Bi$_2$Se$_3$ and Bi$_2$Te$_3$ slabs
are very small (less than 1~\% of 
the squared matrix elements of $\vecop{v}^{\fr}$).

\begin{figure}
\begin{center}
\includegraphics[width=\columnwidth]{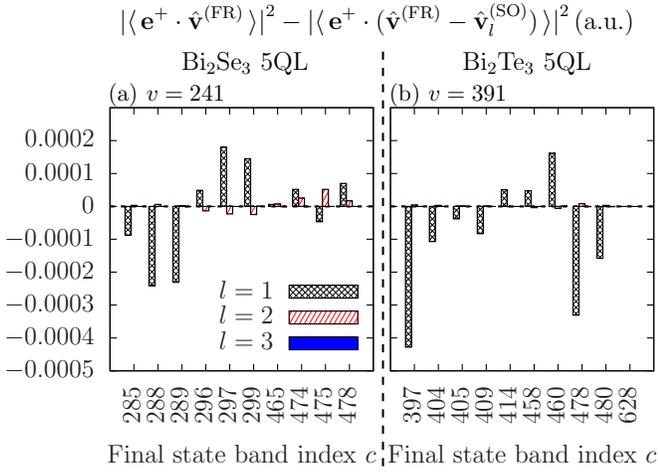}
\end{center}
\caption{
The difference between 
the squared matrix elements
of 5-quintuple-layer slabs of Bi$_2$Se$_3$ and Bi$_2$Te$_3$ 
obtained 
by using the fully-relativistic velocity operator and 
$\hat{\mathbf{v}}^{\mathrm{(FR)}} - \hat{\mathbf{v}}^{\mathrm{(SO)}}_{l}$ [see Eqs.~(\ref{eqn:vfr}) and (\ref{eqn:vso})] 
for the optical transitions 
having the topological surface state with momentum $\mathbf{k}=0.05 \Gamma$K as the initial state.
In all cases, 
left-circularly polarized light was considered.
}
\label{fig:ti_vmat_ldep}
\end{figure}

Figure~\ref{fig:ti_vmat_ldep} shows 
the difference between 
the squared matrix elements of 
$\vecop{v}^{\fr}$ and $\vecop{v}^{\fr} - \vecop{v}^{\so}_{l}$ 
for the same optical transitions presented in Fig.~\ref{fig:ti_vmat}.
We find that 
the effects of the $p$-orbital part 
$\vecop{v}^{\so}_{l=1}$ 
on the optical matrix elements 
are usually 
the largest and 
the effects 
of the $d$- and $f$-orbital parts, 
$\vecop{v}^{\so}_{l=2}$ and $\vecop{v}^{\so}_{l=3}$, 
are much smaller.
This is because 
($i$) the $p$-orbital component of 
the SOC potential $V^{\mathrm{SO}}_{l=1}(r)$
of Bi
is much larger than the $d$- and $f$-orbital components
$V^{\mathrm{SO}}_{l=2, 3}(r)$
(see Figs.~\ref{fig:atom_pseudo} and \ref{fig:atom_vso_integ})
and ($ii$) in particular,
the surface states of Bi$_2$Se$_3$ and Bi$_2$Te$_3$ 
mostly consist of 
the 6$p$ orbitals of Bi atoms.

\begin{figure}
\begin{center}
\includegraphics[width=\columnwidth]{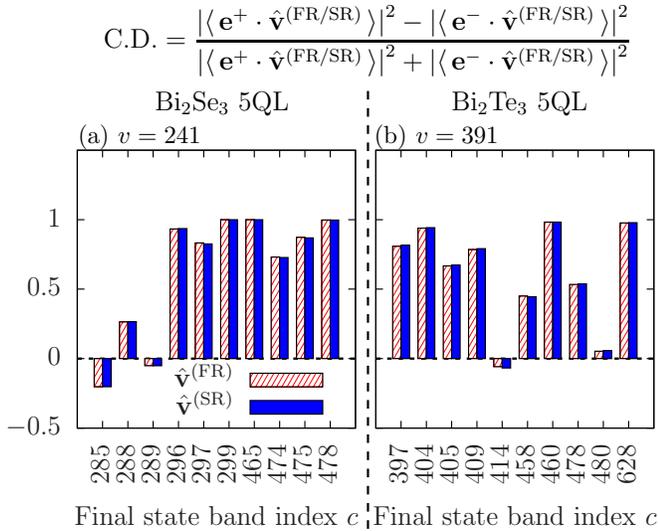}
\end{center}
\caption{
The circular dichroism
of 5-quintuple-layer slabs of 
Bi$_2$Se$_3$ and Bi$_2$Te$_3$
for the optical transitions  
having the topological surface state with momentum $\mathbf{k}=0.05\Gamma$K as the initial state.
}
\label{fig:ti_cd}
\end{figure}

The circular dichroism is defined as 
the relative difference between 
the squared optical matrix elements 
for left- and right-circularly-polarized light 
(see the top of Fig.~\ref{fig:ti_cd}).
We calculated the circular dichroism 
by using $\vecop{v}^{\fr}$ and $\vecop{v}^{\sr}$
and investigated whether 
the effects of the commutator arising from SOC 
change the circular dichroism of 
Bi$_2$Se$_3$ and Bi$_2$Te$_3$ slabs.

In Fig.~\ref{fig:ti_cd}, 
we see that 
the difference between 
the circular dichroisms obtained 
by using $\hat{\mathbf{v}}^{\fr}$ and 
$\hat{\mathbf{v}}^{\sr}$ is negligible.
Contrary to the arguments 
in Ref.~\onlinecite{PhysRevLett.115.016801}, 
the effects of the commutator arising from SOC 
cannot change 
the circular dichroism of 
Bi$_2$Se$_3$ and Bi$_2$Te$_3$ slabs, 
which means that the methods used in 
Refs.~\onlinecite{PhysRevLett.110.216401,PhysRevLett.112.076802,Cao2013} 
will give correct results.
Although we 
did not find the correct final states
(satisfying the proper boundary condition)
in the calculations of 
the optical matrix elements,
because the effects of SOC
on the optical matrix elements
are negligible
over a wide range of energies
($E_{c, \mathbf{k}'} - E_{v, \mathbf{k}'} < 1$~Ry),
it is likely that
imposing the correct boundary condition 
on the final states 
will not make a significant difference
between the results obtained by using 
$\hat{\mathbf{v}}^{\fr}$ and $\hat{\mathbf{v}}^{\sr}$.

\section{Conclusion}

In this study, 
we investigated 
the effects of spin-orbit coupling on 
the optical responses of isolated atoms, 
monolayer transition metal dichaocogenides,
and the topological surface states of  topological insulators 
using first-principles calculations with fully-relativistic pseudopotentials.
By using a method that can separate 
a fully-relativistic Kleinman-Bylander pseudopotential 
into the scalar-relativistic and spin-orbit coupling parts, 
we were able to 
study the effects of spin-orbit coupling on 
the velocity operator and 
its matrix elements
in various systems.

In the case of W and Bi atoms, 
we find that 
the relative contribution of 
the commutator arising from SOC 
to the squared optical matrix elements can be 
4.3~\% for W and 14~\% for Bi.
We find that
the $p$-orbital part of 
the commutator arising from SOC 
gives the largest contribution 
to the optical matrix elements.
The influence of
the $p$-orbital part of 
the spin-orbit coupling potential 
is much larger than 
those of the $d$- and $f$-orbital parts.

In the case of monolayer transition metal dichalcogenides,
the effects of 
the commutator arising from spin-orbit coupling
are much smaller than in the case of atomic systems,
less than 1~\% of
the squared optical matrix elements
for
the optical transitions from 
the valence band edge states.

In the case of five-quintuple layer slabs of 
Bi$_2$Se$_3$ and Bi$_2$Te$_3$,
the effects of spin-orbit coupling on 
the optical matrix elements are again very small
as in the case of 
monolayer transition metal dichalcogenides.
We find that 
the non-local effects of spin-orbit coupling
on the optical matrix elements
are so small that
the effects do not change 
the circular dichroism of 
Bi$_2$Se$_3$ and Bi$_2$Te$_3$ slabs.

In conclusion, 
we confirm that while 
the effects of the commutator arising from spin-orbit coupling
on the optical matrix elements
are not negligible in atomic systems,
the effects are much suppressed
in the cases of monolayer transition metal dichalcogenides 
and topological insulators where the effects of spin-orbit coupling on the electronic structure are considered to be important.

Our calculation results show that 
in studying the optical response of a material 
with heavy elements, 
it is sufficient to calculate 
the optical matrix elements neglecting 
the commutator arising from spin-orbit coupling 
in the velocity operator 
if one has obtained well 
the electronic structure
of the system,
i.e. the energy eigenvalues and eigenstates,
from fully-relativistic first-principles calculations.

T.Y.K. and C.-H.P. were supported by Korean NRF No-2016R1A1A1A05919979 and
by the Creative-Pioneering Research Program through Seoul National University.
AF acknowledges financial support from the EU Centre of Excellence ``MaX - Materials Design at the Exascale'' (Horizon 2020 EINFRA-5, Grant No. 676598).
Computational resources were provided by KISTI Supercomputing Center (KSC-2018-C2-0002).

\bibliography{manuscript}

\begin{thebibliography}{33}%
\makeatletter
\providecommand \@ifxundefined [1]{%
 \@ifx{#1\undefined}
}%
\providecommand \@ifnum [1]{%
 \ifnum #1\expandafter \@firstoftwo
 \else \expandafter \@secondoftwo
 \fi
}%
\providecommand \@ifx [1]{%
 \ifx #1\expandafter \@firstoftwo
 \else \expandafter \@secondoftwo
 \fi
}%
\providecommand \natexlab [1]{#1}%
\providecommand \enquote  [1]{``#1''}%
\providecommand \bibnamefont  [1]{#1}%
\providecommand \bibfnamefont [1]{#1}%
\providecommand \citenamefont [1]{#1}%
\providecommand \href@noop [0]{\@secondoftwo}%
\providecommand \href [0]{\begingroup \@sanitize@url \@href}%
\providecommand \@href[1]{\@@startlink{#1}\@@href}%
\providecommand \@@href[1]{\endgroup#1\@@endlink}%
\providecommand \@sanitize@url [0]{\catcode `\\12\catcode `\$12\catcode
  `\&12\catcode `\#12\catcode `\^12\catcode `\_12\catcode `\%12\relax}%
\providecommand \@@startlink[1]{}%
\providecommand \@@endlink[0]{}%
\providecommand \url  [0]{\begingroup\@sanitize@url \@url }%
\providecommand \@url [1]{\endgroup\@href {#1}{\urlprefix }}%
\providecommand \urlprefix  [0]{URL }%
\providecommand \Eprint [0]{\href }%
\providecommand \doibase [0]{http://dx.doi.org/}%
\providecommand \selectlanguage [0]{\@gobble}%
\providecommand \bibinfo  [0]{\@secondoftwo}%
\providecommand \bibfield  [0]{\@secondoftwo}%
\providecommand \translation [1]{[#1]}%
\providecommand \BibitemOpen [0]{}%
\providecommand \bibitemStop [0]{}%
\providecommand \bibitemNoStop [0]{.\EOS\space}%
\providecommand \EOS [0]{\spacefactor3000\relax}%
\providecommand \BibitemShut  [1]{\csname bibitem#1\endcsname}%
\let\auto@bib@innerbib\@empty
\bibitem [{\citenamefont {Jozwiak}\ \emph {et~al.}(2013)\citenamefont
  {Jozwiak}, \citenamefont {Park}, \citenamefont {Gotlieb}, \citenamefont
  {Hwang}, \citenamefont {Lee}, \citenamefont {Louie}, \citenamefont
  {Denlinger}, \citenamefont {Rotundu}, \citenamefont {Birgeneau},
  \citenamefont {Hussain},\ and\ \citenamefont
  {Lanzara}}]{jozwiak2013photoelectron}%
  \BibitemOpen
  \bibfield  {author} {\bibinfo {author} {\bibfnamefont {C.}~\bibnamefont
  {Jozwiak}}, \bibinfo {author} {\bibfnamefont {C.-H.}\ \bibnamefont {Park}},
  \bibinfo {author} {\bibfnamefont {K.}~\bibnamefont {Gotlieb}}, \bibinfo
  {author} {\bibfnamefont {C.}~\bibnamefont {Hwang}}, \bibinfo {author}
  {\bibfnamefont {D.-H.}\ \bibnamefont {Lee}}, \bibinfo {author} {\bibfnamefont
  {S.~G.}\ \bibnamefont {Louie}}, \bibinfo {author} {\bibfnamefont {J.~D.}\
  \bibnamefont {Denlinger}}, \bibinfo {author} {\bibfnamefont {C.~R.}\
  \bibnamefont {Rotundu}}, \bibinfo {author} {\bibfnamefont {R.~J.}\
  \bibnamefont {Birgeneau}}, \bibinfo {author} {\bibfnamefont {Z.}~\bibnamefont
  {Hussain}}, \ and\ \bibinfo {author} {\bibfnamefont {A.}~\bibnamefont
  {Lanzara}},\ }\bibfield  {title} {\enquote {\bibinfo {title} {Photoelectron
  spin-flipping and texture manipulation in a topological insulator},}\ }\href
  {http://dx.doi.org/10.1038/nphys2572} {\bibfield  {journal} {\bibinfo
  {journal} {Nature Phys.}\ }\textbf {\bibinfo {volume} {9}},\ \bibinfo {pages}
  {293} (\bibinfo {year} {2013})}\BibitemShut {NoStop}%
\bibitem [{\citenamefont {Park}\ and\ \citenamefont
  {Louie}(2012)}]{park2012spin}%
  \BibitemOpen
  \bibfield  {author} {\bibinfo {author} {\bibfnamefont {C.-H.}\ \bibnamefont
  {Park}}\ and\ \bibinfo {author} {\bibfnamefont {S.~G}\ \bibnamefont
  {Louie}},\ }\bibfield  {title} {\enquote {\bibinfo {title} {Spin polarization
  of photoelectrons from topological insulators},}\ }\href@noop {} {\bibfield
  {journal} {\bibinfo  {journal} {Phys. Rev. Lett.}\ }\textbf {\bibinfo
  {volume} {109}},\ \bibinfo {pages} {097601} (\bibinfo {year}
  {2012})}\BibitemShut {NoStop}%
\bibitem [{\citenamefont {Ryoo}\ and\ \citenamefont
  {Park}(2016)}]{ryoo2016spin}%
  \BibitemOpen
  \bibfield  {author} {\bibinfo {author} {\bibfnamefont {J.~H.}\ \bibnamefont
  {Ryoo}}\ and\ \bibinfo {author} {\bibfnamefont {C.-H.}\ \bibnamefont
  {Park}},\ }\bibfield  {title} {\enquote {\bibinfo {title} {Spin-conserving
  and reversing photoemission from the surface states of {B}i$_2${S}e$_3$ and
  {A}u (111)},}\ }\href@noop {} {\bibfield  {journal} {\bibinfo  {journal}
  {Phys. Rev. B}\ }\textbf {\bibinfo {volume} {93}},\ \bibinfo {pages} {085419}
  (\bibinfo {year} {2016})}\BibitemShut {NoStop}%
\bibitem [{\citenamefont {Ryoo}\ and\ \citenamefont {Park}()}]{ryoo2018spin}%
  \BibitemOpen
  \bibfield  {author} {\bibinfo {author} {\bibfnamefont {J.~H.}\ \bibnamefont
  {Ryoo}}\ and\ \bibinfo {author} {\bibfnamefont {C.-H.}\ \bibnamefont
  {Park}},\ }\bibfield  {title} {\enquote {\bibinfo {title} {Momentum-dependent
  spin selection rule in photoemission with glide symmetry},}\ }\href@noop {}
  {\bibinfo  {journal} {arXiv:1807.02368}\ }\BibitemShut {NoStop}%
\bibitem [{\citenamefont {Zhu}\ \emph {et~al.}(2013)\citenamefont {Zhu},
  \citenamefont {Veenstra}, \citenamefont {Levy}, \citenamefont {Ubaldini},
  \citenamefont {Syers}, \citenamefont {Butch}, \citenamefont {Paglione},
  \citenamefont {Haverkort}, \citenamefont {Elfimov},\ and\ \citenamefont
  {Damascelli}}]{PhysRevLett.110.216401}%
  \BibitemOpen
\bibfield  {journal} {  }\bibfield  {author} {\bibinfo {author} {\bibfnamefont
  {Z.-H.}\ \bibnamefont {Zhu}}, \bibinfo {author} {\bibfnamefont {C.~N.}\
  \bibnamefont {Veenstra}}, \bibinfo {author} {\bibfnamefont {G.}~\bibnamefont
  {Levy}}, \bibinfo {author} {\bibfnamefont {A.}~\bibnamefont {Ubaldini}},
  \bibinfo {author} {\bibfnamefont {P.}~\bibnamefont {Syers}}, \bibinfo
  {author} {\bibfnamefont {N.~P.}\ \bibnamefont {Butch}}, \bibinfo {author}
  {\bibfnamefont {J.}~\bibnamefont {Paglione}}, \bibinfo {author}
  {\bibfnamefont {M.~W.}\ \bibnamefont {Haverkort}}, \bibinfo {author}
  {\bibfnamefont {I.~S.}\ \bibnamefont {Elfimov}}, \ and\ \bibinfo {author}
  {\bibfnamefont {A.}~\bibnamefont {Damascelli}},\ }\bibfield  {title}
  {\enquote {\bibinfo {title} {Layer-by-layer entangled spin-orbital texture of
  the topological surface state in {B}i$_2${S}e$_3$},}\ }\href {\doibase
  10.1103/PhysRevLett.110.216401} {\bibfield  {journal} {\bibinfo  {journal}
  {Phys. Rev. Lett.}\ }\textbf {\bibinfo {volume} {110}},\ \bibinfo {pages}
  {216401} (\bibinfo {year} {2013})}\BibitemShut {NoStop}%
\bibitem [{\citenamefont {Zhu}\ \emph {et~al.}(2014)\citenamefont {Zhu},
  \citenamefont {Veenstra}, \citenamefont {Zhdanovich}, \citenamefont
  {Schneider}, \citenamefont {Okuda}, \citenamefont {Miyamoto}, \citenamefont
  {Zhu}, \citenamefont {Namatame}, \citenamefont {Taniguchi}, \citenamefont
  {Haverkort}, \citenamefont {Elfimov},\ and\ \citenamefont
  {Damascelli}}]{PhysRevLett.112.076802}%
  \BibitemOpen
  \bibfield  {author} {\bibinfo {author} {\bibfnamefont {Z.-H.}\ \bibnamefont
  {Zhu}}, \bibinfo {author} {\bibfnamefont {C.~N.}\ \bibnamefont {Veenstra}},
  \bibinfo {author} {\bibfnamefont {S.}~\bibnamefont {Zhdanovich}}, \bibinfo
  {author} {\bibfnamefont {M.~P.}\ \bibnamefont {Schneider}}, \bibinfo {author}
  {\bibfnamefont {T.}~\bibnamefont {Okuda}}, \bibinfo {author} {\bibfnamefont
  {K.}~\bibnamefont {Miyamoto}}, \bibinfo {author} {\bibfnamefont {S.-Y.}\
  \bibnamefont {Zhu}}, \bibinfo {author} {\bibfnamefont {H.}~\bibnamefont
  {Namatame}}, \bibinfo {author} {\bibfnamefont {M.}~\bibnamefont {Taniguchi}},
  \bibinfo {author} {\bibfnamefont {M.~W.}\ \bibnamefont {Haverkort}}, \bibinfo
  {author} {\bibfnamefont {I.~S.}\ \bibnamefont {Elfimov}}, \ and\ \bibinfo
  {author} {\bibfnamefont {A.}~\bibnamefont {Damascelli}},\ }\bibfield  {title}
  {\enquote {\bibinfo {title} {Photoelectron spin-polarization control in the
  topological insulator {B}i$_2${S}e$_3$},}\ }\href {\doibase
  10.1103/PhysRevLett.112.076802} {\bibfield  {journal} {\bibinfo  {journal}
  {Phys. Rev. Lett.}\ }\textbf {\bibinfo {volume} {112}},\ \bibinfo {pages}
  {076802} (\bibinfo {year} {2014})}\BibitemShut {NoStop}%
\bibitem [{\citenamefont {Adolph}\ \emph {et~al.}(1996)\citenamefont {Adolph},
  \citenamefont {Gavrilenko}, \citenamefont {Tenelsen}, \citenamefont
  {Bechstedt},\ and\ \citenamefont {Del~Sole}}]{PhysRevB.53.9797}%
  \BibitemOpen
  \bibfield  {author} {\bibinfo {author} {\bibfnamefont {B.}~\bibnamefont
  {Adolph}}, \bibinfo {author} {\bibfnamefont {V.~I.}\ \bibnamefont
  {Gavrilenko}}, \bibinfo {author} {\bibfnamefont {K.}~\bibnamefont
  {Tenelsen}}, \bibinfo {author} {\bibfnamefont {F.}~\bibnamefont {Bechstedt}},
  \ and\ \bibinfo {author} {\bibfnamefont {R.}~\bibnamefont {Del~Sole}},\
  }\bibfield  {title} {\enquote {\bibinfo {title} {Nonlocality and many-body
  effects in the optical properties of semiconductors},}\ }\href {\doibase
  10.1103/PhysRevB.53.9797} {\bibfield  {journal} {\bibinfo  {journal} {Phys.
  Rev. B}\ }\textbf {\bibinfo {volume} {53}},\ \bibinfo {pages} {9797}
  (\bibinfo {year} {1996})}\BibitemShut {NoStop}%
\bibitem [{\citenamefont {Read}\ and\ \citenamefont
  {Needs}(1991)}]{PhysRevB.44.13071}%
  \BibitemOpen
  \bibfield  {author} {\bibinfo {author} {\bibfnamefont {A.~J.}\ \bibnamefont
  {Read}}\ and\ \bibinfo {author} {\bibfnamefont {R.~J.}\ \bibnamefont
  {Needs}},\ }\bibfield  {title} {\enquote {\bibinfo {title} {Calculation of
  optical matrix elements with nonlocal pseudopotentials},}\ }\href {\doibase
  10.1103/PhysRevB.44.13071} {\bibfield  {journal} {\bibinfo  {journal} {Phys.
  Rev. B}\ }\textbf {\bibinfo {volume} {44}},\ \bibinfo {pages} {13071}
  (\bibinfo {year} {1991})}\BibitemShut {NoStop}%
\bibitem [{\citenamefont {Baroni}\ and\ \citenamefont
  {Resta}(1986)}]{PhysRevB.33.7017}%
  \BibitemOpen
  \bibfield  {author} {\bibinfo {author} {\bibfnamefont {S.}~\bibnamefont
  {Baroni}}\ and\ \bibinfo {author} {\bibfnamefont {R.}~\bibnamefont {Resta}},\
  }\bibfield  {title} {\enquote {\bibinfo {title} {\textit{Ab initio}
  calculation of the macroscopic dielectric constant in silicon},}\ }\href
  {\doibase 10.1103/PhysRevB.33.7017} {\bibfield  {journal} {\bibinfo
  {journal} {Phys. Rev. B}\ }\textbf {\bibinfo {volume} {33}},\ \bibinfo
  {pages} {7017} (\bibinfo {year} {1986})}\BibitemShut {NoStop}%
\bibitem [{\citenamefont {Marini}\ \emph {et~al.}(2001)\citenamefont {Marini},
  \citenamefont {Onida},\ and\ \citenamefont {Del~Sole}}]{PhysRevB.64.195125}%
  \BibitemOpen
  \bibfield  {author} {\bibinfo {author} {\bibfnamefont {A.}~\bibnamefont
  {Marini}}, \bibinfo {author} {\bibfnamefont {G.}~\bibnamefont {Onida}}, \
  and\ \bibinfo {author} {\bibfnamefont {R.}~\bibnamefont {Del~Sole}},\
  }\bibfield  {title} {\enquote {\bibinfo {title} {Plane-wave {DFT}-{LDA}
  calculation of the electronic structure and absorption spectrum of copper},}\
  }\href {\doibase 10.1103/PhysRevB.64.195125} {\bibfield  {journal} {\bibinfo
  {journal} {Phys. Rev. B}\ }\textbf {\bibinfo {volume} {64}},\ \bibinfo
  {pages} {195125} (\bibinfo {year} {2001})}\BibitemShut {NoStop}%
\bibitem [{\citenamefont {Cao}\ \emph {et~al.}(2013)\citenamefont {Cao},
  \citenamefont {Waugh}, \citenamefont {Zhang}, \citenamefont {Luo},
  \citenamefont {Wang}, \citenamefont {Reber}, \citenamefont {Mo},
  \citenamefont {Xu}, \citenamefont {Yang}, \citenamefont {Schneeloch},
  \citenamefont {Gu}, \citenamefont {Brahlek}, \citenamefont {Bansal},
  \citenamefont {Oh}, \citenamefont {Zunger},\ and\ \citenamefont
  {Dessau}}]{Cao2013}%
  \BibitemOpen
  \bibfield  {author} {\bibinfo {author} {\bibfnamefont {Y.}~\bibnamefont
  {Cao}}, \bibinfo {author} {\bibfnamefont {J.~A.}\ \bibnamefont {Waugh}},
  \bibinfo {author} {\bibfnamefont {X.-W.}\ \bibnamefont {Zhang}}, \bibinfo
  {author} {\bibfnamefont {J.-W.}\ \bibnamefont {Luo}}, \bibinfo {author}
  {\bibfnamefont {Q.}~\bibnamefont {Wang}}, \bibinfo {author} {\bibfnamefont
  {T.~J.}\ \bibnamefont {Reber}}, \bibinfo {author} {\bibfnamefont {S.~K.}\
  \bibnamefont {Mo}}, \bibinfo {author} {\bibfnamefont {Z.}~\bibnamefont {Xu}},
  \bibinfo {author} {\bibfnamefont {A.}~\bibnamefont {Yang}}, \bibinfo {author}
  {\bibfnamefont {J.}~\bibnamefont {Schneeloch}}, \bibinfo {author}
  {\bibfnamefont {G.~D.}\ \bibnamefont {Gu}}, \bibinfo {author} {\bibfnamefont
  {M.}~\bibnamefont {Brahlek}}, \bibinfo {author} {\bibfnamefont
  {N.}~\bibnamefont {Bansal}}, \bibinfo {author} {\bibfnamefont
  {S.}~\bibnamefont {Oh}}, \bibinfo {author} {\bibfnamefont {A.}~\bibnamefont
  {Zunger}}, \ and\ \bibinfo {author} {\bibfnamefont {D.~S.}\ \bibnamefont
  {Dessau}},\ }\bibfield  {title} {\enquote {\bibinfo {title} {Mapping the
  orbital wavefunction of the surface states in three-dimensional topological
  insulators},}\ }\href {http://dx.doi.org/10.1038/nphys2685} {\bibfield
  {journal} {\bibinfo  {journal} {Nature Phys.}\ }\textbf {\bibinfo {volume}
  {9}},\ \bibinfo {pages} {499} (\bibinfo {year} {2013})}\BibitemShut {NoStop}%
\bibitem [{\citenamefont {Xu}\ \emph {et~al.}(2015)\citenamefont {Xu},
  \citenamefont {Liu}, \citenamefont {Yukawa}, \citenamefont {Zhang},
  \citenamefont {Matsuda}, \citenamefont {Miller},\ and\ \citenamefont
  {Chiang}}]{PhysRevLett.115.016801}%
  \BibitemOpen
  \bibfield  {author} {\bibinfo {author} {\bibfnamefont {C.-Z.}\ \bibnamefont
  {Xu}}, \bibinfo {author} {\bibfnamefont {Y.}~\bibnamefont {Liu}}, \bibinfo
  {author} {\bibfnamefont {R.}~\bibnamefont {Yukawa}}, \bibinfo {author}
  {\bibfnamefont {L.-X.}\ \bibnamefont {Zhang}}, \bibinfo {author}
  {\bibfnamefont {I.}~\bibnamefont {Matsuda}}, \bibinfo {author} {\bibfnamefont
  {T.}~\bibnamefont {Miller}}, \ and\ \bibinfo {author} {\bibfnamefont {T.-C.}\
  \bibnamefont {Chiang}},\ }\bibfield  {title} {\enquote {\bibinfo {title}
  {Photoemission circular dichroism and spin polarization of the topological
  surface states in ultrathin {B}i$_2${T}e$_3$ films},}\ }\href {\doibase
  10.1103/PhysRevLett.115.016801} {\bibfield  {journal} {\bibinfo  {journal}
  {Phys. Rev. Lett.}\ }\textbf {\bibinfo {volume} {115}},\ \bibinfo {pages}
  {016801} (\bibinfo {year} {2015})}\BibitemShut {NoStop}%
\bibitem [{\citenamefont {Kleinman}(1980)}]{PhysRevB.21.2630}%
  \BibitemOpen
  \bibfield  {author} {\bibinfo {author} {\bibfnamefont {L.}~\bibnamefont
  {Kleinman}},\ }\bibfield  {title} {\enquote {\bibinfo {title} {Relativistic
  norm-conserving pseudopotential},}\ }\href {\doibase
  10.1103/PhysRevB.21.2630} {\bibfield  {journal} {\bibinfo  {journal} {Phys.
  Rev. B}\ }\textbf {\bibinfo {volume} {21}},\ \bibinfo {pages} {2630}
  (\bibinfo {year} {1980})}\BibitemShut {NoStop}%
\bibitem [{\citenamefont {Sakurai}\ and\ \citenamefont
  {Napolitano}(2011)}]{sakurai2011modern}%
  \BibitemOpen
  \bibfield  {author} {\bibinfo {author} {\bibfnamefont {J.~J.}\ \bibnamefont
  {Sakurai}}\ and\ \bibinfo {author} {\bibfnamefont {J.}~\bibnamefont
  {Napolitano}},\ }\href@noop {} {\emph {\bibinfo {title} {Modern quantum
  mechanics}}}\ (\bibinfo  {publisher} {Addison-Wesley},\ \bibinfo {year}
  {2011})\BibitemShut {NoStop}%
\bibitem [{\citenamefont {Bachelet}\ \emph {et~al.}(1982)\citenamefont
  {Bachelet}, \citenamefont {Hamann},\ and\ \citenamefont
  {Schl\"uter}}]{PhysRevB.26.4199}%
  \BibitemOpen
  \bibfield  {author} {\bibinfo {author} {\bibfnamefont {G.~B.}\ \bibnamefont
  {Bachelet}}, \bibinfo {author} {\bibfnamefont {D.~R.}\ \bibnamefont
  {Hamann}}, \ and\ \bibinfo {author} {\bibfnamefont {M.}~\bibnamefont
  {Schl\"uter}},\ }\bibfield  {title} {\enquote {\bibinfo {title}
  {Pseudopotentials that work: From {H} to {P}u},}\ }\href {\doibase
  10.1103/PhysRevB.26.4199} {\bibfield  {journal} {\bibinfo  {journal} {Phys.
  Rev. B}\ }\textbf {\bibinfo {volume} {26}},\ \bibinfo {pages} {4199}
  (\bibinfo {year} {1982})}\BibitemShut {NoStop}%
\bibitem [{\citenamefont {Schiff}(1968)}]{schiff1968quantum}%
  \BibitemOpen
  \bibfield  {author} {\bibinfo {author} {\bibfnamefont {L.~I.}\ \bibnamefont
  {Schiff}},\ }\href@noop {} {\emph {\bibinfo {title} {Quantum Mechanics}}}\
  (\bibinfo  {publisher} {McGraw-Hill},\ \bibinfo {address} {New York},\
  \bibinfo {year} {1968})\BibitemShut {NoStop}%
\bibitem [{\citenamefont {Hybertsen}\ and\ \citenamefont
  {Louie}(1986)}]{PhysRevB.34.2920}%
  \BibitemOpen
  \bibfield  {author} {\bibinfo {author} {\bibfnamefont {M.~S.}\ \bibnamefont
  {Hybertsen}}\ and\ \bibinfo {author} {\bibfnamefont {S.~G.}\ \bibnamefont
  {Louie}},\ }\bibfield  {title} {\enquote {\bibinfo {title} {Spin-orbit
  splitting in semiconductors and insulators from the \textit{ab initio}
  pseudopotential},}\ }\href {\doibase 10.1103/PhysRevB.34.2920} {\bibfield
  {journal} {\bibinfo  {journal} {Phys. Rev. B}\ }\textbf {\bibinfo {volume}
  {34}},\ \bibinfo {pages} {2920} (\bibinfo {year} {1986})}\BibitemShut
  {NoStop}%
\bibitem [{\citenamefont {Kleinman}\ and\ \citenamefont
  {Bylander}(1982)}]{PhysRevLett.48.1425}%
  \BibitemOpen
  \bibfield  {author} {\bibinfo {author} {\bibfnamefont {L.}~\bibnamefont
  {Kleinman}}\ and\ \bibinfo {author} {\bibfnamefont {D.~M.}\ \bibnamefont
  {Bylander}},\ }\bibfield  {title} {\enquote {\bibinfo {title} {Efficacious
  form for model pseudopotentials},}\ }\href {\doibase
  10.1103/PhysRevLett.48.1425} {\bibfield  {journal} {\bibinfo  {journal}
  {Phys. Rev. Lett.}\ }\textbf {\bibinfo {volume} {48}},\ \bibinfo {pages}
  {1425} (\bibinfo {year} {1982})}\BibitemShut {NoStop}%
\bibitem [{\citenamefont {Hemstreet}\ \emph {et~al.}(1993)\citenamefont
  {Hemstreet}, \citenamefont {Fong},\ and\ \citenamefont
  {Nelson}}]{PhysRevB.47.4238}%
  \BibitemOpen
  \bibfield  {author} {\bibinfo {author} {\bibfnamefont {L.~A.}\ \bibnamefont
  {Hemstreet}}, \bibinfo {author} {\bibfnamefont {C.~Y.}\ \bibnamefont {Fong}},
  \ and\ \bibinfo {author} {\bibfnamefont {J.~S.}\ \bibnamefont {Nelson}},\
  }\bibfield  {title} {\enquote {\bibinfo {title} {First-principles
  calculations of spin-orbit splittings in solids using nonlocal separable
  pseudopotentials},}\ }\href {\doibase 10.1103/PhysRevB.47.4238} {\bibfield
  {journal} {\bibinfo  {journal} {Phys. Rev. B}\ }\textbf {\bibinfo {volume}
  {47}},\ \bibinfo {pages} {4238} (\bibinfo {year} {1993})}\BibitemShut
  {NoStop}%
\bibitem [{\citenamefont {Theurich}\ and\ \citenamefont
  {Hill}(2001)}]{PhysRevB.64.073106}%
  \BibitemOpen
  \bibfield  {author} {\bibinfo {author} {\bibfnamefont {G.}~\bibnamefont
  {Theurich}}\ and\ \bibinfo {author} {\bibfnamefont {N.~A.}\ \bibnamefont
  {Hill}},\ }\bibfield  {title} {\enquote {\bibinfo {title} {Self-consistent
  treatment of spin-orbit coupling in solids using relativistic fully separable
  \textit{ab initio} pseudopotentials},}\ }\href {\doibase
  10.1103/PhysRevB.64.073106} {\bibfield  {journal} {\bibinfo  {journal} {Phys.
  Rev. B}\ }\textbf {\bibinfo {volume} {64}},\ \bibinfo {pages} {073106}
  (\bibinfo {year} {2001})}\BibitemShut {NoStop}%
\bibitem [{\citenamefont {Perdew}\ \emph {et~al.}(1996)\citenamefont {Perdew},
  \citenamefont {Burke},\ and\ \citenamefont
  {Ernzerhof}}]{PhysRevLett.77.3865}%
  \BibitemOpen
  \bibfield  {author} {\bibinfo {author} {\bibfnamefont {J.~P.}\ \bibnamefont
  {Perdew}}, \bibinfo {author} {\bibfnamefont {K.}~\bibnamefont {Burke}}, \
  and\ \bibinfo {author} {\bibfnamefont {M.}~\bibnamefont {Ernzerhof}},\
  }\bibfield  {title} {\enquote {\bibinfo {title} {Generalized gradient
  approximation made simple},}\ }\href {\doibase 10.1103/PhysRevLett.77.3865}
  {\bibfield  {journal} {\bibinfo  {journal} {Phys. Rev. Lett.}\ }\textbf
  {\bibinfo {volume} {77}},\ \bibinfo {pages} {3865} (\bibinfo {year}
  {1996})}\BibitemShut {NoStop}%
\bibitem [{\citenamefont {Giannozzi}\ \emph {et~al.}(2009)\citenamefont
  {Giannozzi}, \citenamefont {Baroni}, \citenamefont {Bonini}, \citenamefont
  {Calandra}, \citenamefont {Car}, \citenamefont {Cavazzoni}, \citenamefont
  {Ceresoli}, \citenamefont {Chiarotti}, \citenamefont {Cococcioni},
  \citenamefont {Dabo}, \citenamefont {Dal~Corso}, \citenamefont {Gironcoli},
  \citenamefont {Fabris}, \citenamefont {Fratesi}, \citenamefont {Gebauer},
  \citenamefont {Gerstmann}, \citenamefont {Gougoussis}, \citenamefont
  {Kokalj}, \citenamefont {Lazzeri}, \citenamefont {Martin-Samos},
  \citenamefont {Marzari}, \citenamefont {Mauri}, \citenamefont {Mazzarello},
  \citenamefont {Paolini}, \citenamefont {Pasquarello}, \citenamefont
  {Paulatto}, \citenamefont {Sbraccia}, \citenamefont {Scandolo}, \citenamefont
  {Sclauzero}, \citenamefont {Seitsonen}, \citenamefont {Smogunov},
  \citenamefont {Umari},\ and\ \citenamefont
  {Wentzcovitch}}]{0953-8984-21-39-395502}%
  \BibitemOpen
  \bibfield  {author} {\bibinfo {author} {\bibfnamefont {P.}~\bibnamefont
  {Giannozzi}}, \bibinfo {author} {\bibfnamefont {S.}~\bibnamefont {Baroni}},
  \bibinfo {author} {\bibfnamefont {N.}~\bibnamefont {Bonini}}, \bibinfo
  {author} {\bibfnamefont {M.}~\bibnamefont {Calandra}}, \bibinfo {author}
  {\bibfnamefont {R.}~\bibnamefont {Car}}, \bibinfo {author} {\bibfnamefont
  {C.}~\bibnamefont {Cavazzoni}}, \bibinfo {author} {\bibfnamefont
  {D.}~\bibnamefont {Ceresoli}}, \bibinfo {author} {\bibfnamefont {G.~L.}\
  \bibnamefont {Chiarotti}}, \bibinfo {author} {\bibfnamefont {M.}~\bibnamefont
  {Cococcioni}}, \bibinfo {author} {\bibfnamefont {I.}~\bibnamefont {Dabo}},
  \bibinfo {author} {\bibfnamefont {A.}~\bibnamefont {Dal~Corso}}, \bibinfo
  {author} {\bibfnamefont {S.~de}\ \bibnamefont {Gironcoli}}, \bibinfo {author}
  {\bibfnamefont {S.}~\bibnamefont {Fabris}}, \bibinfo {author} {\bibfnamefont
  {G.}~\bibnamefont {Fratesi}}, \bibinfo {author} {\bibfnamefont
  {R.}~\bibnamefont {Gebauer}}, \bibinfo {author} {\bibfnamefont
  {U.}~\bibnamefont {Gerstmann}}, \bibinfo {author} {\bibfnamefont
  {C.}~\bibnamefont {Gougoussis}}, \bibinfo {author} {\bibfnamefont
  {A.}~\bibnamefont {Kokalj}}, \bibinfo {author} {\bibfnamefont
  {M.}~\bibnamefont {Lazzeri}}, \bibinfo {author} {\bibfnamefont
  {L.}~\bibnamefont {Martin-Samos}}, \bibinfo {author} {\bibfnamefont
  {N.}~\bibnamefont {Marzari}}, \bibinfo {author} {\bibfnamefont
  {F.}~\bibnamefont {Mauri}}, \bibinfo {author} {\bibfnamefont
  {R.}~\bibnamefont {Mazzarello}}, \bibinfo {author} {\bibfnamefont
  {S.}~\bibnamefont {Paolini}}, \bibinfo {author} {\bibfnamefont
  {A.}~\bibnamefont {Pasquarello}}, \bibinfo {author} {\bibfnamefont
  {L.}~\bibnamefont {Paulatto}}, \bibinfo {author} {\bibfnamefont
  {C.}~\bibnamefont {Sbraccia}}, \bibinfo {author} {\bibfnamefont
  {S.}~\bibnamefont {Scandolo}}, \bibinfo {author} {\bibfnamefont
  {G.}~\bibnamefont {Sclauzero}}, \bibinfo {author} {\bibfnamefont {A.~P.}\
  \bibnamefont {Seitsonen}}, \bibinfo {author} {\bibfnamefont {A.}~\bibnamefont
  {Smogunov}}, \bibinfo {author} {\bibfnamefont {P.}~\bibnamefont {Umari}}, \
  and\ \bibinfo {author} {\bibfnamefont {R.~M.}\ \bibnamefont {Wentzcovitch}},\
  }\bibfield  {title} {\enquote {\bibinfo {title} {{QUANTUM ESPRESSO}: a
  modular and open-source software project for quantum simulations of
  materials},}\ }\href {http://stacks.iop.org/0953-8984/21/i=39/a=395502}
  {\bibfield  {journal} {\bibinfo  {journal} {J. Phys.: Condens. Matter}\
  }\textbf {\bibinfo {volume} {21}},\ \bibinfo {pages} {395502} (\bibinfo
  {year} {2009})}\BibitemShut {NoStop}%
\bibitem [{\citenamefont {Giannozzi}\ \emph {et~al.}(2017)\citenamefont
  {Giannozzi}, \citenamefont {Andreussi}, \citenamefont {Brumme}, \citenamefont
  {Bunau}, \citenamefont {{Buongiorno Nardelli}}, \citenamefont {Calandra},
  \citenamefont {Car}, \citenamefont {Cavazzoni}, \citenamefont {Ceresoli},
  \citenamefont {Cococcioni}, \citenamefont {Colonna}, \citenamefont
  {Carnimeo}, \citenamefont {{Dal Corso}}, \citenamefont {{de Gironcoli}},
  \citenamefont {Delugas}, \citenamefont {{DiStasio Jr.}}, \citenamefont
  {Ferretti}, \citenamefont {Floris}, \citenamefont {Fratesi}, \citenamefont
  {Fugallo}, \citenamefont {Gebauer}, \citenamefont {Gerstmann}, \citenamefont
  {Giustino}, \citenamefont {Gorni}, \citenamefont {Jia}, \citenamefont
  {Kawamura}, \citenamefont {Ko}, \citenamefont {Kokalj}, \citenamefont
  {K\"u\c{c}\"ukbenli}, \citenamefont {Lazzeri}, \citenamefont {Marsili},
  \citenamefont {Marzari}, \citenamefont {Mauri}, \citenamefont {Nguyen},
  \citenamefont {Nguyen}, \citenamefont {de-la Roza}, \citenamefont {Paulatto},
  \citenamefont {Ponc\'e}, \citenamefont {Rocca}, \citenamefont {Sabatini},
  \citenamefont {Santra}, \citenamefont {Schlipf}, \citenamefont {Seitsonen},
  \citenamefont {Smogunov}, \citenamefont {Timrov}, \citenamefont {Thonhauser},
  \citenamefont {Umari}, \citenamefont {Vast}, \citenamefont {Wu},\ and\
  \citenamefont {Baroni}}]{giannozzi2017jpcm}%
  \BibitemOpen
  \bibfield  {author} {\bibinfo {author} {\bibfnamefont {P.}~\bibnamefont
  {Giannozzi}}, \bibinfo {author} {\bibfnamefont {O.}~\bibnamefont
  {Andreussi}}, \bibinfo {author} {\bibfnamefont {T.}~\bibnamefont {Brumme}},
  \bibinfo {author} {\bibfnamefont {O.}~\bibnamefont {Bunau}}, \bibinfo
  {author} {\bibfnamefont {M.}~\bibnamefont {{Buongiorno Nardelli}}}, \bibinfo
  {author} {\bibfnamefont {M.}~\bibnamefont {Calandra}}, \bibinfo {author}
  {\bibfnamefont {R.}~\bibnamefont {Car}}, \bibinfo {author} {\bibfnamefont
  {C.}~\bibnamefont {Cavazzoni}}, \bibinfo {author} {\bibfnamefont
  {D.}~\bibnamefont {Ceresoli}}, \bibinfo {author} {\bibfnamefont
  {M.}~\bibnamefont {Cococcioni}}, \bibinfo {author} {\bibfnamefont
  {N.}~\bibnamefont {Colonna}}, \bibinfo {author} {\bibfnamefont
  {I.}~\bibnamefont {Carnimeo}}, \bibinfo {author} {\bibfnamefont
  {A.}~\bibnamefont {{Dal Corso}}}, \bibinfo {author} {\bibfnamefont
  {S.}~\bibnamefont {{de Gironcoli}}}, \bibinfo {author} {\bibfnamefont
  {P.}~\bibnamefont {Delugas}}, \bibinfo {author} {\bibfnamefont {R.~A.}\
  \bibnamefont {{DiStasio Jr.}}}, \bibinfo {author} {\bibfnamefont
  {A.}~\bibnamefont {Ferretti}}, \bibinfo {author} {\bibfnamefont
  {A.}~\bibnamefont {Floris}}, \bibinfo {author} {\bibfnamefont
  {G.}~\bibnamefont {Fratesi}}, \bibinfo {author} {\bibfnamefont
  {G.}~\bibnamefont {Fugallo}}, \bibinfo {author} {\bibfnamefont
  {R.}~\bibnamefont {Gebauer}}, \bibinfo {author} {\bibfnamefont
  {U.}~\bibnamefont {Gerstmann}}, \bibinfo {author} {\bibfnamefont
  {F.}~\bibnamefont {Giustino}}, \bibinfo {author} {\bibfnamefont
  {T.}~\bibnamefont {Gorni}}, \bibinfo {author} {\bibfnamefont
  {J.}~\bibnamefont {Jia}}, \bibinfo {author} {\bibfnamefont {M.}~\bibnamefont
  {Kawamura}}, \bibinfo {author} {\bibfnamefont {H.-Y.}\ \bibnamefont {Ko}},
  \bibinfo {author} {\bibfnamefont {A.}~\bibnamefont {Kokalj}}, \bibinfo
  {author} {\bibfnamefont {E.}~\bibnamefont {K\"u\c{c}\"ukbenli}}, \bibinfo
  {author} {\bibfnamefont {M.}~\bibnamefont {Lazzeri}}, \bibinfo {author}
  {\bibfnamefont {M.}~\bibnamefont {Marsili}}, \bibinfo {author} {\bibfnamefont
  {N.}~\bibnamefont {Marzari}}, \bibinfo {author} {\bibfnamefont
  {F.}~\bibnamefont {Mauri}}, \bibinfo {author} {\bibfnamefont {N.~L.}\
  \bibnamefont {Nguyen}}, \bibinfo {author} {\bibfnamefont {H.-V.}\
  \bibnamefont {Nguyen}}, \bibinfo {author} {\bibfnamefont {A.~Otero}\
  \bibnamefont {de-la Roza}}, \bibinfo {author} {\bibfnamefont
  {L.}~\bibnamefont {Paulatto}}, \bibinfo {author} {\bibfnamefont
  {S.}~\bibnamefont {Ponc\'e}}, \bibinfo {author} {\bibfnamefont
  {D.}~\bibnamefont {Rocca}}, \bibinfo {author} {\bibfnamefont
  {R.}~\bibnamefont {Sabatini}}, \bibinfo {author} {\bibfnamefont
  {B.}~\bibnamefont {Santra}}, \bibinfo {author} {\bibfnamefont
  {M.}~\bibnamefont {Schlipf}}, \bibinfo {author} {\bibfnamefont {A.~P.}\
  \bibnamefont {Seitsonen}}, \bibinfo {author} {\bibfnamefont {A.}~\bibnamefont
  {Smogunov}}, \bibinfo {author} {\bibfnamefont {I.}~\bibnamefont {Timrov}},
  \bibinfo {author} {\bibfnamefont {T.}~\bibnamefont {Thonhauser}}, \bibinfo
  {author} {\bibfnamefont {P.}~\bibnamefont {Umari}}, \bibinfo {author}
  {\bibfnamefont {N.}~\bibnamefont {Vast}}, \bibinfo {author} {\bibfnamefont
  {X.}~\bibnamefont {Wu}}, \ and\ \bibinfo {author} {\bibfnamefont
  {S.}~\bibnamefont {Baroni}},\ }\bibfield  {title} {\enquote {\bibinfo {title}
  {Advanced capabilities for materials modelling with {Q}uantum {ESPRESSO}},}\
  }\href@noop {} {\bibfield  {journal} {\bibinfo  {journal} {J. Phys.: Condens.
  Matter}\ }\textbf {\bibinfo {volume} {29}},\ \bibinfo {pages} {465901}
  (\bibinfo {year} {2017})}\BibitemShut {NoStop}%
\bibitem [{\citenamefont {Marini}\ \emph {et~al.}(2009)\citenamefont {Marini},
  \citenamefont {Hogan}, \citenamefont {Gr\"uning},\ and\ \citenamefont
  {Varsano}}]{Marini20091392}%
  \BibitemOpen
  \bibfield  {author} {\bibinfo {author} {\bibfnamefont {A.}~\bibnamefont
  {Marini}}, \bibinfo {author} {\bibfnamefont {C.}~\bibnamefont {Hogan}},
  \bibinfo {author} {\bibfnamefont {M.}~\bibnamefont {Gr\"uning}}, \ and\
  \bibinfo {author} {\bibfnamefont {Daniele}\ \bibnamefont {Varsano}},\
  }\bibfield  {title} {\enquote {\bibinfo {title} {yambo: An \textit{ab initio}
  tool for excited state calculations},}\ }\href {\doibase
  http://dx.doi.org/10.1016/j.cpc.2009.02.003} {\bibfield  {journal} {\bibinfo
  {journal} {Comput. Phys. Commun.}\ }\textbf {\bibinfo {volume} {180}},\
  \bibinfo {pages} {1392} (\bibinfo {year} {2009})}\BibitemShut {NoStop}%
\bibitem [{\citenamefont {Hamann}(2013)}]{PhysRevB.88.085117}%
  \BibitemOpen
  \bibfield  {author} {\bibinfo {author} {\bibfnamefont {D.~R.}\ \bibnamefont
  {Hamann}},\ }\bibfield  {title} {\enquote {\bibinfo {title} {Optimized
  norm-conserving {V}anderbilt pseudopotentials},}\ }\href {\doibase
  10.1103/PhysRevB.88.085117} {\bibfield  {journal} {\bibinfo  {journal} {Phys.
  Rev. B}\ }\textbf {\bibinfo {volume} {88}},\ \bibinfo {pages} {085117}
  (\bibinfo {year} {2013})}\BibitemShut {NoStop}%
\bibitem [{\citenamefont {Schlipf}\ and\ \citenamefont
  {F.}(2015)}]{Schlipf201536}%
  \BibitemOpen
  \bibfield  {author} {\bibinfo {author} {\bibfnamefont {M.}~\bibnamefont
  {Schlipf}}\ and\ \bibinfo {author} {\bibfnamefont {Gygi}\ \bibnamefont
  {F.}},\ }\bibfield  {title} {\enquote {\bibinfo {title} {Optimization
  algorithm for the generation of {ONCV} pseudopotentials},}\ }\href {\doibase
  http://dx.doi.org/10.1016/j.cpc.2015.05.011} {\bibfield  {journal} {\bibinfo
  {journal} {Comput. Phys. Commun.}\ }\textbf {\bibinfo {volume} {196}},\
  \bibinfo {pages} {36} (\bibinfo {year} {2015})}\BibitemShut {NoStop}%
\bibitem [{\citenamefont {Burke}\ and\ \citenamefont
  {Grant}(1967)}]{burke1967effect}%
  \BibitemOpen
  \bibfield  {author} {\bibinfo {author} {\bibfnamefont {V.~M.}\ \bibnamefont
  {Burke}}\ and\ \bibinfo {author} {\bibfnamefont {I.~P.}\ \bibnamefont
  {Grant}},\ }\bibfield  {title} {\enquote {\bibinfo {title} {The effect of
  relativity on atomic wave functions},}\ }\href@noop {} {\bibfield  {journal}
  {\bibinfo  {journal} {Proc. Phys. Soc.}\ }\textbf {\bibinfo {volume} {90}},\
  \bibinfo {pages} {297} (\bibinfo {year} {1967})}\BibitemShut {NoStop}%
\bibitem [{\citenamefont {Griffiths}(2005)}]{griffiths2005introduction}%
  \BibitemOpen
  \bibfield  {author} {\bibinfo {author} {\bibfnamefont {D.~J.}\ \bibnamefont
  {Griffiths}},\ }\href@noop {} {\emph {\bibinfo {title} {Introduction to
  quantum mechanics}}}\ (\bibinfo  {publisher} {Pearson Prentice Hall},\
  \bibinfo {address} {Upper Saddle River, NJ},\ \bibinfo {year}
  {2005})\BibitemShut {NoStop}%
\bibitem [{Note1()}]{Note1}%
  \BibitemOpen
  \bibinfo {note} {See, for example, Fig.~6.9 of Ref.~\protect \rev@citealpnum
  {griffiths2005introduction}}\BibitemShut {NoStop}%
\bibitem [{\citenamefont {Monkhorst}\ and\ \citenamefont
  {Pack}(1976)}]{PhysRevB.13.5188}%
  \BibitemOpen
  \bibfield  {author} {\bibinfo {author} {\bibfnamefont {H.~J.}\ \bibnamefont
  {Monkhorst}}\ and\ \bibinfo {author} {\bibfnamefont {J.~D.}\ \bibnamefont
  {Pack}},\ }\bibfield  {title} {\enquote {\bibinfo {title} {Special points for
  {B}rillouin-zone integrations},}\ }\href {\doibase 10.1103/PhysRevB.13.5188}
  {\bibfield  {journal} {\bibinfo  {journal} {Phys. Rev. B}\ }\textbf {\bibinfo
  {volume} {13}},\ \bibinfo {pages} {5188} (\bibinfo {year}
  {1976})}\BibitemShut {NoStop}%
\bibitem [{\citenamefont {Zhu}\ \emph {et~al.}(2011)\citenamefont {Zhu},
  \citenamefont {Cheng},\ and\ \citenamefont
  {Schwingenschl\"ogl}}]{PhysRevB.84.153402}%
  \BibitemOpen
  \bibfield  {author} {\bibinfo {author} {\bibfnamefont {Z.~Y.}\ \bibnamefont
  {Zhu}}, \bibinfo {author} {\bibfnamefont {Y.~C.}\ \bibnamefont {Cheng}}, \
  and\ \bibinfo {author} {\bibfnamefont {U.}~\bibnamefont
  {Schwingenschl\"ogl}},\ }\bibfield  {title} {\enquote {\bibinfo {title}
  {Giant spin-orbit-induced spin splitting in two-dimensional transition-metal
  dichalcogenide semiconductors},}\ }\href {\doibase
  10.1103/PhysRevB.84.153402} {\bibfield  {journal} {\bibinfo  {journal} {Phys.
  Rev. B}\ }\textbf {\bibinfo {volume} {84}},\ \bibinfo {pages} {153402}
  (\bibinfo {year} {2011})}\BibitemShut {NoStop}%
\bibitem [{\citenamefont {Gibertini}\ \emph {et~al.}(2014)\citenamefont
  {Gibertini}, \citenamefont {Pellegrino}, \citenamefont {Marzari},\ and\
  \citenamefont {Polini}}]{PhysRevB.90.245411}%
  \BibitemOpen
  \bibfield  {author} {\bibinfo {author} {\bibfnamefont {M.}~\bibnamefont
  {Gibertini}}, \bibinfo {author} {\bibfnamefont {F.~M.~D.}\ \bibnamefont
  {Pellegrino}}, \bibinfo {author} {\bibfnamefont {N.}~\bibnamefont {Marzari}},
  \ and\ \bibinfo {author} {\bibfnamefont {M.}~\bibnamefont {Polini}},\
  }\bibfield  {title} {\enquote {\bibinfo {title} {Spin-resolved optical
  conductivity of two-dimensional group-{VIB} transition-metal
  dichalcogenides},}\ }\href {\doibase 10.1103/PhysRevB.90.245411} {\bibfield
  {journal} {\bibinfo  {journal} {Phys. Rev. B}\ }\textbf {\bibinfo {volume}
  {90}},\ \bibinfo {pages} {245411} (\bibinfo {year} {2014})}\BibitemShut
  {NoStop}%
\bibitem [{\citenamefont {Cao}\ \emph {et~al.}(2012)\citenamefont {Cao},
  \citenamefont {Wang}, \citenamefont {Han}, \citenamefont {Ye}, \citenamefont
  {Zhu}, \citenamefont {Shi}, \citenamefont {Niu}, \citenamefont {Tan},
  \citenamefont {Wang}, \citenamefont {Liu},\ and\ \citenamefont
  {Feng}}]{Cao2012}%
  \BibitemOpen
  \bibfield  {author} {\bibinfo {author} {\bibfnamefont {T.}~\bibnamefont
  {Cao}}, \bibinfo {author} {\bibfnamefont {G.}~\bibnamefont {Wang}}, \bibinfo
  {author} {\bibfnamefont {W.}~\bibnamefont {Han}}, \bibinfo {author}
  {\bibfnamefont {H.}~\bibnamefont {Ye}}, \bibinfo {author} {\bibfnamefont
  {C.}~\bibnamefont {Zhu}}, \bibinfo {author} {\bibfnamefont {J.}~\bibnamefont
  {Shi}}, \bibinfo {author} {\bibfnamefont {Q.}~\bibnamefont {Niu}}, \bibinfo
  {author} {\bibfnamefont {P.}~\bibnamefont {Tan}}, \bibinfo {author}
  {\bibfnamefont {E.}~\bibnamefont {Wang}}, \bibinfo {author} {\bibfnamefont
  {B.}~\bibnamefont {Liu}}, \ and\ \bibinfo {author} {\bibfnamefont
  {J.}~\bibnamefont {Feng}},\ }\bibfield  {title} {\enquote {\bibinfo {title}
  {Valley-selective circular dichroism of monolayer molybdenum disulphide},}\
  }\href {\doibase 10.1038/ncomms1882} {\bibfield  {journal} {\bibinfo
  {journal} {Nature Commun.}\ }\textbf {\bibinfo {volume} {3}},\ \bibinfo
  {pages} {887} (\bibinfo {year} {2012})}\BibitemShut {NoStop}%
\end{thebibliography}%

\end{document}